\documentclass[galaxies,review,accept,pdftex,moreauthors]{Definitions/mdpi} 

\firstpage{1} 
\pubvolume{1}
\issuenum{1}
\articlenumber{0}
\pubyear{2022}
\copyrightyear{2022}
\externaleditor{Academic Editor: {Giovanni De Cesare}} 
\datereceived{1 April 2022} 
\dateaccepted{24 May 2022} 
\datepublished{30 May 2022} 
\hreflink{https://doi.org/} 
\pdfoutput=1

\newcommand{\be}{\begin{equation}}
\newcommand{\ee}{\end{equation}}
\newcommand{\ba}{\begin{eqnarray}}
\newcommand{\ea}{\end{eqnarray}}

\newcommand{\fracb}[2]{\left(\frac{#1}{#2}\right)}

\newcommand{\mean}[1]{\langle{#1}\rangle}

\usepackage{enumitem}
\usepackage{aas_macros}
\usepackage{amssymb}
\usepackage{amsmath}
\usepackage{subfig}

\definecolor{blazeorange}{rgb}{1.0, 0.4, 0.0}
\definecolor{seagreen}{rgb}{0.18, 0.55, 0.34}
\definecolor{rufous}{rgb}{0.66, 0.11, 0.03}
\definecolor{royalfuchsia}{rgb}{0.79, 0.17, 0.57}
\definecolor{scarlet}{rgb}{1.0, 0.13, 0.0}
\definecolor{royalpurple}{rgb}{0.47, 0.32, 0.66}

\Title{Gamma-Ray Bursts at TeV Energies: Theoretical Considerations}

\TitleCitation{Gamma-Ray Bursts at TeV Energies: Theoretical Considerations}

\Author{{Ramandeep} 
 Gill $^{1,2,3,}$*\orcidA{} and Jonathan Granot $^{2,3,4,}$*\orcidB{}}

\AuthorNames{Ramandeep Gill and Jonathan Granot}

\AuthorCitation{Gill, R.; Granot, J.}

\address{%
$^{1}$ \quad Instituto de Radioastronomía y Astrofísica, Universidad Nacional Autónoma de México, Antigua Carretera a Pátzcuaro \# 8701, 
Ex-Hda. San José de la Huerta, {Morelia 58089}
, Michoacán, Mexico\\
$^{2}$ \quad Astrophysics Research Center {of} 
 the Open University (ARCO), The Open University of Israel, {P.O. Box 808,} 
 Ra'anana 4353701, Israel\\
$^{3}$ \quad Department of Physics, The George Washington University, Washington, DC 20052, USA\\
$^{4}$ \quad Department of Natural Sciences, The Open University of Israel, {P.O. Box 808}, Ra'anana 4353701, Israel
}
\corres{Correspondence: r.gill@irya.unam.mx (R.G.); granot@openu.ac.il (J.G.)}

\abstract{Gamma-ray bursts (GRBs) are the most luminous explosions in the Universe and are powered by ultra-relativistic jets. 
Their prompt $\gamma$-ray emission briefly outshines the rest of the $\gamma$-ray sky, making them detectable from cosmological 
distances. A burst is followed by, and sometimes partially overlaps with, a similarly energetic but very broadband and longer-lasting 
afterglow emission. While most GRBs are detected below a few MeV, over 100 have been detected at high ($\gtrsim$0.1~GeV) energies, 
and several have now been observed up to tens of GeV with the \textit{Fermi} Large Area Telescope (LAT). A new electromagnetic 
window in the very-high-energy (VHE) domain ($\gtrsim$0.1~TeV) was recently opened with the detection of an afterglow emission in the 
$(0.1$\textendash$1)\,$TeV energy band by ground-based imaging atmospheric Cherenkov telescopes. The 
emission mechanism for the VHE spectral component is not fully understood, and its detection offers important 
constraints for GRB physics. This review provides a brief overview of the different leptonic and hadronic mechanisms capable 
of producing a VHE emission in GRBs. The same mechanisms possibly give rise to the high-energy spectral component seen during 
the prompt emission of many \textit{Fermi}-LAT GRBs. Possible origins of its delayed onset and long duration well into 
the afterglow phase, with implications for the emission region and relativistic collisionless shock physics, are discussed. 
Key results for using GRBs as ideal probes for constraining models of extra-galactic background light and intergalactic magnetic 
fields, as well as for testing Lorentz invariance violation, are presented.
}


\begin{document}
\section{Introduction}

Gamma-ray bursts (GRBs) are cataclysmic events that occur at cosmological distances. 
(See, e.g., \cite{Piran-04,Zhang-Meszaros-04,Meszaros-06,Kumar-Zhang-15}~for a comprehensive review.) 
They are the most electromagnetically luminous transient phenomena in the Universe. GRBs 
involve the explosive release of energy over a short timescale, producing a burst 
of $\gamma$-rays with isotropic-equivalent luminosity of 
$L_{\gamma,\rm iso}$$\sim$10$^{51}$--10$^{54}$~${\rm erg\,s}^{-1}$. Their emission is powered 
by ultrarelativistic (with bulk Lorentz factors $\Gamma\gtrsim100$) 
bipolar collimated outflows driven by a compact object central engine. The~identity of the central 
engine, which could be either a black hole (BH) or a millisecond magnetar, is not entirely clear as the highly variable emission is produced far away from it at a 
radial distance of $R\sim10^{12}$--$10^{16}$~cm. The~most luminous phase of the burst, referred to as the ``prompt'' phase, is short-lived with a bimodal duration 
distribution separated at $t\sim 2\,$s, where the short and long GRBs have typical observed durations of $t_{\rm GRB}\sim10^{-0.5}\,$s and $t_{\rm GRB}\sim10^{1.5}\,$s, 
respectively \citep{Kouveliotou+93}. These two classes of GRBs are also distinct spectrally, with~the short GRBs being spectrally harder compared to the long GRBs that produce softer $\gamma$-rays. 
The long-soft GRBs are associated with the core collapse 
of massive ($\gtrsim$(20--30)$M_\odot$) Wolf--Rayet stars \citep{Woosley-93,WB06}, whereas (at least some) 
short-hard GRBs originate in compact object mergers of two neutron stars (NSs) or a NS-BH binary \citep{Eichler+89,Narayan+92}. The~first-ever detection of 
a short GRB coincident with gravitational waves (GWs) from the merger of two NSs came from GW 170817/GRB 170817A 
\citep{Abbott+17-GW170817-Ligo-Detection,Abbott+17-GW170817-GRB170817A}.

Many details of the prompt GRB emission, in~particular, the~energy dissipation process, the~exact radiation mechanism, and~the transfer of radiation in 
the highly dynamical flow remain poorly understood. All of these different processes combine to produce a non-thermal spectrum that is often well-described 
by the Band-function \citep{Band+93}, an~empirical fit to the spectrum featuring a smoothly broken power law. This break manifests as 
a peak in the $\nu F_\nu$ spectrum, at~a mean photon energy $\langle E_{\rm br}\rangle\simeq250\,$keV, with~the asymptotic power-law photon indices below and above the peak energy having 
mean values of $\langle\alpha_{\rm Band}\rangle\simeq-1$ and $\langle\beta_{\rm Band}\rangle\simeq-2.3$, respectively \citep{Preece+00,Kaneko+06}. 

While most of the energy in the prompt GRB comes out at $E\simeq E_{\rm br}$, the~featureless power-law spectrum above this energy 
extends beyond $100\,$MeV in most GRBs detected by the \textit{Fermi}-Large Area Telescope (LAT) \citep{Ajello+19}, with~a high-energy 
spectral cutoff seen in only about 20\% of the cases, e.g.,~\cite{Ackermann+12}, which are most likely caused by intrinsic 
opacity to pair production \citep{Ackermann+12,Vianello+18}. In~rare cases, the~prompt GRB spectrum shows an additional hard spectral 
component that extends beyond $\sim$10~GeV, as~seen by the \textit{Fermi}-LAT, and~well into very high energies ($\gtrsim$0.1~TeV), 
as seen by ground-based atmospheric Cherenkov telescopes, e.g.,~MAGIC and H.E.S.S. (See, e.g., \cite{Nava-21,Noda-Parsons-22}~for a review.) 
This high-energy (HE; $\gtrsim$100~MeV) emission overlaps with the sub-MeV prompt GRB, and both the HE and very- high-energy (VHE) 
emissions persist throughout the afterglow phase---the much longer-lasting and broadband (X-rays/Optical/Radio) emission that follows 
the short-lived prompt phase. The~spectral and temporal properties of the HE emission provide a glimpse into the global energetics of 
the bursts as well as yield important constraints on GRB physics that cannot be obtained from the sub-MeV emission~alone.

The main objective of this review is to provide a concise summary of the widely discussed radiation mechanisms that may explain the 
spectral and temporal properties of the VHE and/or HE emission in GRBs. We first discuss several HE/VHE radiation mechanisms 
in Section~\ref{sec:HE-Emission-Mechanisms} and provide some of the fundamental quantities that can be calculated and compared to observations. 
This is followed by a discussion of the delayed HE emission, additional prompt GRB spectral component at high energies, and~long-lived HE emission 
seen by \textit{Fermi}-LAT as well as popular theoretical explanations offered for it, along with implications for the bulk Lorentz factor 
$\Gamma$, in~Section~\ref{sec:Prompt-HE-Emission}. 
Next, Section~\ref{sec:GeV-Afterglow-GRB130427A} presents an overview of the HE afterglow seen in the exceptionally bright GRB 130427A, along with 
several important implications for the radiation mechanism and relativistic shock acceleration physics. The~recent detection of a 
$\sim$TeV afterglow emission by MAGIC and H.E.S.S in only a few GRBs and key implications of such a detection for GRB physics are 
discussed in Section~\ref{sec:TeV-Afterglow}. The~use of HE photons from distant GRBs as a probe of extra-galactic background light (EBL), 
inter-galactic magnetic field, and~Lorentz invariance violation are the topics of discussion in Section~\ref{sec:Non-GRB-Physics}. Finally, 
in Section~\ref{sec:Outstanding-Qs}, we end this review with important outstanding questions in GRB physics and present closing remarks in 
Section~\ref{sec:Closing-Remarks}.
 
\section{Relevant High-Energy or Very-High-Energy Emission~Mechanisms}
\label{sec:HE-Emission-Mechanisms}

There are several HE/VHE $\gamma$-ray emission mechanisms that operate wherever particles (leptons and hadrons) are accelerated to or 
generated with high Lorentz factors (LFs). In~GRBs, the emission regions can be either internal to the relativistic outflow, e.g.,~at 
internal shocks or magnetic reconnection sites, or~external to it, e.g.,~in the shocked external medium behind the external forward 
(afterglow) shock, or~even at larger distances from the outflow. Below,~we review some of the widely discussed processes that are capable 
of producing HE to VHE $\gamma$-ray photons. Other more detailed reviews on this topic are \citep{Fan-Piran-08,Kumar-Zhang-15}.

\subsection{Electron Synchrotron~Emission}

Relativistic electrons with LFs $\gamma_e\gg1$ cool by emitting synchrotron photons when gyrating around magnetic field lines with 
comoving magnetic field strength $B'$ (all primed quantities are in the comoving/fluid rest frame). At~collisionless shocks (internal 
or external), a~fraction $\xi_e$ of the electrons are accelerated into a non-thermal power-law energy distribution, 
$dN/d\gamma_e\propto\gamma_e^{-p}$ for $\gamma_m\leq\gamma_e\leq\gamma_M$ and $2\lesssim p\lesssim3$, that holds a fraction $\epsilon_e$ 
of the post-shock internal energy density, and~arises due to Fermi acceleration 
\citep{Fermi-49,Axford+77,Bell-78}. The~minimal LF of this distribution is
\begin{equation}\label{eq:gamma_m}
    \gamma_m=\frac{\epsilon_e}{\xi_e}\fracb{p-2}{p-1}\frac{m_p}{m_e}(\Gamma_{ud}-1)
\end{equation}
where $\Gamma_{ud}$ is the relative LF between the regions upstream and downstream of the shock front. The~resulting (observed) 
\textit{optically thin} synchrotron spectrum in this case comprises multiple power-law 
segments joined smoothly at characteristic 
break energies \citep{Sari+98,Granot-Sari-02} (shown here for fiducial parameters relevant for prompt emission for which 
$\Gamma\gg1$ and $\beta\simeq1$),
\begin{eqnarray}
    E_m &=& \frac{\Gamma}{(1+z)}h\nu_m' = \frac{\Gamma}{(1+z)}\gamma_m^2\frac{\hbar eB'}{m_ec} 
    \simeq \frac{664}{(1+z)}\frac{\gamma_{m,3.5}^2f_{\sigma,-2}^{1/2}L_{\rm iso,52}^{1/2}}{R_{13}}\,{\rm keV} \\
    E_c &=& \frac{\Gamma}{(1+z)}h\nu_c' = \frac{36\pi^2}{(1+z)}\frac{\hbar em_ec^3}{\sigma_T^2}\frac{\Gamma^3\beta^2}{B'^3R^2} 
    \simeq \frac{3.3\times10^{-4}}{(1+z)}\frac{\Gamma_{2.5}^6R_{13}}{f_{\sigma,-2}^{3/2}L_{\rm iso,52}^{3/2}}\,{\rm keV}\,,
\end{eqnarray}
where $\hbar=h/2\pi$ with $h$ being the Planck's constant, $\sigma_T$ is the Thomson cross section, $e$ is the elementary charge, $m_e$ is the electron rest mass, 
and $c$ is the speed of light. The~energy $E_m$ corresponds to the characteristic synchrotron frequency ($\nu_m'$) of minimal energy electrons with LF $\gamma_m$, 
and the cooling break energy $E_c$ corresponds to the cooling frequency ($\nu_c'$) of electrons with LF 
$\gamma_c = (6\pi m_ec^2/\sigma_T)(\Gamma\beta/B'^2R)\approx2.2R_{13}\Gamma_{2.5}^3f_{\sigma,-2}^{-1}L_{\rm iso,52}^{-1}$ 
that are cooling at the dynamical time, such that their synchrotron cooling time, $t_{\rm syn}'=6\pi m_ec/\sigma_TB'^2\gamma_e$, equals the dynamical time, 
$t_{\rm cool}'=t_{\rm dyn}'=R/\Gamma\beta c$. For~some model parameters, $\gamma_c<1$, which is obviously unphysical, but~instead represents very rapid cooling 
of particles to non-relativistic velocities in less than the dynamical time \citep{Guetta-Granot-03a}. As~a result, relativistically hot particles only occupy a thin 
layer behind the shock which is a fraction $\gamma_c$ of the comoving width $\Delta'$ of the ejecta shell, where the electrons are cold in the remaining majority 
of the shell. 
In the above equations, we have expressed the comoving magnetic field in terms of the more useful quantities, using the fact that the total isotropic-equivalent 
power of the outflow can be written in terms of $L_{\rm k,iso}$ and $L_{B,\rm iso}$, the~kinetic energy and magnetic field powers, respectively. As a result, 
 
$L_{\rm iso}=L_{\gamma,\rm iso}/\epsilon_\gamma=L_{\rm k,iso}+L_{\rm B,iso}=L_{B,\rm iso}/f_\sigma=R^2\Gamma^2\beta cB'^2/f_\sigma$, 
where $f_\sigma=\sigma/(1+\sigma)$ is the fraction of total power carried by the magnetic field with $\sigma=L_{\rm B,iso}/L_{\rm k,iso}$ being the outflow magnetization, 
and $L_{\gamma,\rm iso}$ is the isotropic-equivalent $\gamma$-ray luminosity which is a fraction $\epsilon_\gamma$ of the total power. This yields the comoving B-field  
strength $B'\approx1.8\times10^4f_{\sigma,-2}^{1/2}L_{\rm iso,52}^{1/2}R_{13}^{-1}\Gamma_{2.5}^{-1}\,$G with $\beta\simeq1$ for an ultra-relativistic flow. 
 The ordering of the break energies depends on whether the electrons are in the \textit{fast cooling} regime, for~
which $E_c < E_m$, or~the \textit{slow cooling} regime, with~$E_m < E_c$. This relative ordering also decides the values of the spectral indices of the flux 
density $F_E$ for the different power law segments,
\begin{equation}
    \frac{d\log F_E}{d\log E} = \begin{cases}
    1/3,\quad\quad & E < \min(E_c,E_m) \\
    -1/2\quad\quad & E_c < E < E_m\quad(\rm fast~cooling) \\
    -(p-1)/2\quad\quad & E_m < E < Ec\quad(\rm slow~cooling)\,. \\
    -p/2,\quad\quad & E > \max(E_c,E_m) \\
    \end{cases}
\end{equation}

The emission in the power-law segment above the spectral peak energy ($\max(E_c,E_m)$) can only extend up to the maximum synchrotron energy $E_{\rm syn,max}$. 
This energy depends on the efficiency of the acceleration process while the charged particles (electrons or protons) lose energy to synchrotron cooling. 
The typical timescale $t'_{\rm acc}$ over which particles, say the electrons with LF $\gamma_e$, are accelerated as they are scattered across the relativistic 
shock is at best the Larmor time $t_L' = \gamma_em_ec/eB'$, i.e.,~$t'_L/t'_{\rm acc}=\kappa_{\rm acc}\leq1$. Their radiative cooling timescale $t'_c$ is at 
most $t_{\rm syn}'$ as any additional radiative cooling besides synchrotron (e.g., inverse-Compton) would only shorten $t'_c$, i.e.,~$t'_c=\kappa_c t'_{\rm syn}$ 
with $\kappa_c\leq1$. Equating the acceleration and radiative cooling timescales, $t'_{\rm acc}=t'_c$, yields the maximum LF attained by the electrons, 
$\gamma_M = (6\pi e\kappa/\sigma_TB')^{1/2}$ where $\kappa=\kappa_{\rm acc}\kappa_c\leq1$. These electrons then radiate at the characteristic synchrotron 
energy, e.g., \cite{Guilbert+83,deJager-Harding-92,Piran-Nakar-10,Kumar+12,Atwood+13}
\begin{equation}
    E_{\rm syn,max} = \frac{\Gamma}{(1+z)}\gamma_M^2\frac{\hbar eB'}{m_ec} 
    = \frac{\Gamma}{(1+z)}\kappa \frac{m_ec^2}{\alpha_F} 
    \simeq \frac{7.0\,\kappa}{(1+z)}\Gamma_2\,{\rm GeV}\,,
\end{equation}
where $\alpha_F=e^2/\hbar c \simeq 1/137$ is the fine structure constant, and $\kappa$ is a factor expected to be of order unity that depends mainly on the 
details of particle acceleration and diffusion in the shock downstream and~upstream.

It is therefore challenging to explain VHE photons as arising from synchrotron emission by electrons. In~addition, 
depending on the compactness of the emission region,  emission can be suppressed due to $e^\pm$-pair production via 
$\gamma\gamma$-annihilation ($\gamma\gamma\to e^-e^+)$,  e.g.,~\citep{Fenimore+93,Woods-Loeb-95,Baring-Harding-97,Lithwick-Sari-01,Granot+08,Gill-Granot-18a}. 
This poses more of a problem for the prompt emission and less so for the afterglow. Alternatively, the~VHE photons can be explained 
by proton synchrotron emission (see Section~\ref{sec:proton-synchro}) or synchrotron self-compton (SSC; see Section~\ref{sec:SSC}) emission 
by the same electron population that produced the seed synchrotron~radiation. 

\subsection{Proton Synchrotron~Emission}\label{sec:proton-synchro}
\textls[-15]{High-energy protons that are accelerated at shocks (like the electrons) to LFs $\gamma_p$ can also cool by emitting 
synchrotron photons in magnetized regions \citep{Bottcher-Dermer-98,Totani-98}. However, the~emitted power per particle 
($P_{\rm syn}\propto\sigma_T\gamma_i^2{B^2}\propto(\gamma_i/m_i)^2{B^2}$ for $i=\{e,p\}$) is much smaller where 
it is suppressed by a factor $(m_e/m_p)^2 \simeq (1836)^{-2} \simeq 3\times10^{-7}$ with respect to that for electrons 
when $\gamma_p=\gamma_e$ (and suppressed by the square of this factor for \mbox{$E_p=E_e\gg m_pc^2$}) since the Thomson 
scattering cross section for protons is much smaller, $\sigma_{T,p}=(m_e/m_p)^2\sigma_T$, than~that of electrons. 
{To compensate for this suppression, the~magnetic field in the emission region must be larger than that obtained 
in a leptonic synchrotron scenario, so much so that the magnetic field energy would hold a good fraction of the total 
energy \citep{Razzaque+10}}. The~characteristic synchrotron energy of minimal energy protons is 
$E_{m,p} = (\gamma_{m,p}/\gamma_{m,e})^2(m_e/m_p)E_m\approx[\xi_e{\epsilon_p}/\xi_p\epsilon_e]^2(m_e/m_p)^3E_m$ (assuming the electrons and protons hold 
fractions $\epsilon_e$ and {$\epsilon_p=1-\epsilon_e-\epsilon_B$} of the post-shock internal energy, 
and that fractions $\xi_e$ and $\xi_p$ of the electrons and protons, respectively, 
form a power-law energy distribution), and~the cooling break energy is $E_{c,p}=(m_p/m_e)^5E_c$, with~the corresponding LF 
$\gamma_{c,p}=(m_p/m_e)^3\gamma_c$. As~a result, the~maximum LF of protons accelerated at the same shock as electrons is $\gamma_{M,p}=(m_p/m_e)\gamma_M$, which 
yields $E_{\rm p,syn,max}\simeq13(1+z)^{-1}\kappa_p\Gamma_2\,$TeV, e.g., \cite{Totani-98}.}

Recent suggestions replacing electron with proton synchrotron emisson have been made 
to explain the apparent low-energy (below the spectral peak) spectral breaks that are difficult to explain with electron synchrotron emission, e.g., \cite{Ghisellini+19}. 
However, knowing that protons are inefficient at radiating away their internal (or random-motion) energy as compared to electrons, the~significant 
reduction in radiative efficiency must be compensated by having a much larger total energy budget, a~requirement that may be too demanding, e.g., \cite{Wang+09,Dermer-13}. 
Moreover, in~such a scenario it would also be very difficult to suppress the much more efficient radiation from the electrons for it to not over-power that from the~protons. 

\subsection{Synchrotron Self-Compton (SSC)}
\label{sec:SSC}
A distribution of relativistic electrons can inverse-Compton scatter some of the same synchrotron photons that it produced, 
leading to a synchrotron self-Compton emission. When the energy of the incoming synchrotron photon in the rest frame of the scattering 
electron is much smaller than the electron's rest energy, $E''_{\rm syn}\sim\gamma_eE'_{\rm syn}\ll m_ec^2$, then 
the scattering occurs in the Thomson regime (where the electron's recoil can be neglected) and is called \textit{elastic} or \textit{coherent}. 
The scattered photon emerges with an energy of 
$E_{\rm SSC}'\sim\gamma_eE''_{\rm SSC}\cong\gamma_eE''_{\rm syn}\sim\gamma_e^2E_{\rm syn}'$. The~additional cooling of particles due to inverse-Compton 
scattering introduces a factor of $(1+Y)$ in the cooling time, such that $t'_c=t'_{\rm syn}/(1+Y)$. 
Here $Y(\gamma_e) \equiv P'_{\rm IC}(\gamma_e)/P'_{\rm syn}(\gamma_e)$ is the Compton-$y$ parameter given by the ratio of the 
power radiated in the IC component to that in the synchrotron~component. 

\textls[-15]{In the Thomson regime, $P_{\{\rm IC,\,\rm syn\}}'= (4/3)\sigma_T c (\gamma_e^2-1)U_{\{\gamma,\,B\}}'$ for isotropic emission, which yields 
$Y=U_\gamma'/U'_B$ (where $U'_B=B'^2/8\pi$) that is independent of $\gamma_e$, where $U_\gamma'$ is the energy density of the seed synchrotron emission that is IC scattered by the electrons. If~this seed radiation arises from shock-heated electrons, 
$U_\gamma'=\eta\beta U_e'/(1+Y)$, e.g., \citep{Sari-Esin-01}, where $\eta=\min[1,(\nu_m/\nu_c)^{(p-2)/2}]$ is the fraction of electron 
energy radiated away in synchrotron and IC photons, and~$\beta$ is the downstream velocity relative to the shock front (and is order 
unity for a relativistic shock). With~$U_e'=\epsilon_e U_{\rm int}'$ 
and $U_B'=\epsilon_B U_{\rm int}'$, where $\epsilon_e$ and $\epsilon_B$ are the fractions of the total internal energy behind the shock 
($U_{\rm int}'$) that goes into accelerating electrons and generating the magnetic fields, the~expression for $Y$ simplifies to, 
\mbox{e.g., \cite{Panaitescu-Kumar-00,Sari-Esin-01,Zhang-Meszaros-01,Guetta-Granot-03a,Nakar+09},}}
\begin{equation}
    Y = \frac{\sqrt{1+4\eta\epsilon_e/\epsilon_B}-1}{2} \approx \begin{cases}
    \eta\epsilon_e/\epsilon_B,\quad\quad\;\, \eta\epsilon_e/\epsilon_B\ll1\,, \\
    \sqrt{\eta\epsilon_e/\epsilon_B},\quad\ \  \eta\epsilon_e/\epsilon_B\gg1\,. \\
    \end{cases}
\end{equation}

When $\eta\epsilon_e\ll\epsilon_B$, then $Y\ll1$, and Compton cooling is negligible. Otherwise, the~extra cooling also means that 
the maximum particle LF is reduced, $\tilde{\gamma}_M=(1+Y)^{-1/2}\gamma_M$, and~likewise $\tilde E_{\rm syn,max}=(1+Y)^{-1}E_{\rm syn,max}$. The~characteristic 
spectral break energies of the SSC spectrum corresponding to that of the synchrotron spectrum are
\begin{eqnarray}
    E_m^{SSC} &\approx& 2\gamma_m^2E_m \approx 
    \frac{4.2}{(1+z)}\frac{\gamma_{m,3.5}^4f_{\sigma,-3}^{1/2}L_{\rm iso,52}^{1/2}}{R_{13}}\,{\rm TeV} \\
    E_c^{SSC} &\approx& 2\tilde\gamma_c^2\tilde E_c 
    \approx \frac{10}{(1+z)(1+Y)^{4}}\frac{\Gamma_{2.5}^{12}R_{13}^3}{f_{\sigma,-3}^{7/2}L_{\rm iso,52}^{7/2}}\,{\rm keV}\,,
\end{eqnarray}
where $\tilde\gamma_c=\gamma_c/(1+Y)$, and $\tilde E_c=E_c/(1+Y)^2$. The~maximum energy of an inverse-Compton scattered photon is 
$E_{\rm IC,max}'=\gamma_em_ec^2$, and~since $\gamma_e\leq\tilde\gamma_M$ for a power-law electron distribution,
\begin{equation}
    E_{\rm max}^{\rm SSC} = \frac{\Gamma}{(1+z)}\tilde\gamma_Mm_ec^2 
    \simeq \frac{250}{(1+z)(1+Y)^{1/2}}\frac{R_{13}^{1/2}\Gamma_{2.5}^{3/2}}{f_{\sigma,-3}^{1/4}L_{\rm iso,52}^{1/4}}\,{\rm TeV}\,. 
\end{equation}

When the energy of the incoming photon in the rest frame of the scattering electron exceeds the rest mass energy of the electron, 
$E''_{\rm syn}\sim\gamma_eE_{\rm syn}'>m_ec^2$, the~recoil suffered by the electron can no longer be ignored, and quantum corrections need to be taken into account. 
The scattering no longer occurs in the Thomson regime, and the correct scattering cross section in this case is the Klien--Nishina cross section ($\sigma_{\rm KN}$), 
which depends on the energy of the incoming photon \citep{Rybicki-Lightman-79}. For~incoming photon energy $x=h\nu''/m_ec^2\gg 1$, the~scattering cross section 
is highly suppressed, with~$\sigma_{\rm KN}(x) \propto x^{-1}$. Moreover, the~electron recoil implies that $E'_{\rm SSC}(\gamma_e)\sim\gamma_em_ec^2=E'_{\rm KN}(\gamma_e)$ 
in this limit. Therefore, IC scattering can efficiently cool an electron with LF $\gamma_e$ only for seed synchrotron photons with energies 
$E'_{\rm syn}<m_ec^2/\gamma_e$. Thus, accounting of Klein--Nishina effects causes the Compton-Y parameter of each electron to depend on its LF, 
$Y=Y(\gamma_e)\equiv P'_{\rm IC}/P'_{\rm syn}\approx U'_\gamma[E'_{\rm syn}<m_ec^2/\gamma_e]/U'_B$. This may cause interesting modifications of the spectrum 
(in both the synchrotron and SSC components) when $\eta\epsilon_e\gg\epsilon_B$ \citep{Nakar+09}. Notice that since $Y=Y(\gamma_e)$ may vary between different 
electrons, it is natural to define the global Compton-Y parameter by $\bar{Y}=L_{\rm IC}/L_{\rm syn}$, which is the mean value of $Y(\gamma_e)$ weighted by the 
synchrotron emissivity. Therefore, the~SSC flux is suppressed above the photon energy
\begin{equation}\label{E_KN}
    E_{\rm KN}=\frac{\Gamma}{(1+z)} E_{\rm KN}' \approx \frac{\Gamma}{(1+z)} \gamma_e m_ec^2 
    = \frac{\Gamma^2}{(1+z)^2} \frac{m_e^2c^4}{\max(E_m,E_c)}\,,
\end{equation}
where $E = E_m$ ($E = E_c)$ are the energies where the synchrotron $\nu F_\nu$ spectrum peaks in the fast (slow) cooling scenario. Likewise, the~
spectral peak of the SSC spectrum occurs at $E = E_m^{\rm SSC} (E = E_c^{\rm SSC})$ in the fast (slow) cooling case. The~ratio of the spectral 
peak flux is given by $E^{\rm SSC}F_{E}^{\rm SSC} / E F_E \approx L_{\rm IC}/L_{\rm syn}\equiv\bar{Y}$. 
If $\max(\gamma_m^3E_m,\tilde\gamma_c^3\tilde E_c) < \Gamma m_ec^2/(1+z)$, second order SSC scatterings also occur in the Thomson regime, and if 
$\bar{Y}$ is not much smaller than unity, a third spectral peak can appear, e.g., \cite{Stern-Poutanen-04,Kobayashi+07}.

\subsection{External~Inverse-Compton}
External inverse-Compton emission (EIC) arises when the softer seed photons are inverse-Compton scattered to high energies by relativistic electrons in a location 
physically distinct from where the seed photons were produced. This can occur in several different ways, e.g.,~(i) seed photons produced in internal 
dissipation and upscattered by forward-shock or reverse-shock-heated electrons 
\citep{Beloborodov-05a,Beloborodov-05b,Wang+06,Murase+10,Murase+11,Beloborodov+14,Murase+18,Theodore+21}, 
(ii) seed photons produced in the reverse shock 
and upscattered by forward shock-heated-electrons \citep{Panaitescu-Meszaros-98b,Wang+01}, (iii) seed photons produced in the forward shock and upscattered 
by reverse shock-heated electrons~\citep{Panaitescu-Meszaros-98b,Wang+01}, (iv) externally produced ambient seed photons, e.g.,~from the accretion disk 
\citep{Shaviv-Dar-95} or the massive star progenitor's envelope \citep{Lazzati+00}, are upscattered by cold electrons in the relativistic outflow in a process 
also referred to as \textit{bulk Compton scattering} or Compton drag, (v) photospheric seed photons in the relativistic baryon-poor 
jet upscattered by the shocked electrons in the shock transition layer between the baryon-poor jet and baryon-loaded envelope \citep{Eichler-Levinson-03}, 
{and (vi) seed photons provided by the cocoon \citep{Kimura+19}, after~it breaks out of the dynamical ejecta in a NS-NS merger, or~that 
from the AGN disk \citep{Yuan+21}, if~the merger occurs inside the disk of an AGN, that are IC upscattered 
to VHE $\gamma$-rays by electrons energized in the dissipation of prolonged jets powered by late-time central engine activity.}
As an illustrative example, below~we summarize the important points for the simplest case in scenario (i) and provide estimates of 
the maximum photon energy obtained in this process when the X-ray flare emission overlaps with the external forward shock electrons 
\citep{Wang+06}.

\subsubsection*{IC Scattering of X-ray Flare Photons by External Forward Shock~Electrons}
As the relativistic ejecta plows through the circumburst medium (CBM), with~density $\rho = AR^{-k}$ where $R$ is the radial distance 
from the central engine, it is slowed down. In~the process, two shocks are formed where the shocked regions are separated by a contact 
discontinuity that has a bulk LF $\Gamma$. The~\textit{forward} shock runs ahead of the contact discontinuity with bulk LF $\Gamma_{\rm fs}=\sqrt{2}\Gamma$, 
sweeping up the CBM and shock-heating it. The~\textit{reverse} shock moves backward (in the rest frame of 
the contact discontinuity) into the ejecta, decelerating and shock-heating it. In~the following, we adopt the \textit{thin-shell} case for which the 
reverse shock is Newtonian (or mildly relativistic). Alternatively, the~reverse shock becomes relativistic before crossing the ejecta shell in the 
\textit{thick-shell} case, which we will not discuss here (but see \citet{Sari-Piran-95}). Most of the isotropic-equivalent kinetic 
energy of the ejecta ($E_{\rm k,iso}$) is transferred to the kinetic and internal energy of the shock-heated swept up CBM 
behind the forward shock at the deceleration radius,
\begin{equation}\label{eq:R,Gamma}
    R_{\rm dec} = \left[\frac{(3-k)E_{\rm k,iso}}{4\pi Ac^2\Gamma_0^2}\right]^{1/(3-k)},\quad\quad
    \Gamma(R) \approx \begin{cases}
    \Gamma_0\, & R \leq R_{\rm dec} \\
    \Gamma_0(R/R_{\rm dec})^{-(3-k)/2}\, & R > R_{\rm dec}
    \end{cases}
\end{equation}
where $A=m_pn=1.67\times10^{-24}n_0\,{\rm g~cm}^{-3}$ for $k=0$ (ISM) and $A=\dot M/4\pi v_w=5\times10^{11}A_\star\,{\rm g~cm}^{-1}$ for $k=2$ 
(wind medium; $A_\star=1$ corresponds to a mass loss rate of $\dot M=10^{-5}M_\odot\,{\rm yr}^{-1}$ with a wind speed of $v_w=10^8\,{\rm cm\,s}^{-1}$), 
and $\Gamma_0\gg1$ is the initial LF of the relativistic ejecta at which it coasts for $R<R_{\rm dec}$. For~$R>R_{\rm dec}$, 
the blast wave dynamics become self-similar, and the bulk LF of the shocked material decays as a power 
law in $R$ \citep{Blandford-McKee-76}. The~transition for $\Gamma(R)$ from the coasting to the self-similar power-law phase is smooth 
in general, but~here we use the broken power-law approximation in Equation~(\ref{eq:R,Gamma}) for~simplicity.

The LF of minimal energy power-law electrons accelerated at collisionless shocks is given by $\gamma_m$ in Equation~(\ref{eq:gamma_m}). 
For electrons accelerated at the forward shock $\Gamma_{ud} = \Gamma(R)\gg1$, in~which case the minimal particle LF for $R>R_{\rm dec}$ 
is given by (for $p=2.5$)
\begin{eqnarray}
    \gamma_{f,m} &\simeq& \frac{2\times10^3}{(1+z)^{-3/8}}\fracb{E_{\rm k,iso,53}}{n_0}^{1/8}\fracb{\epsilon_{e,-1}}{\xi_e}t_3^{-3/8} 
    \quad\quad(k=0) \\
    &\simeq& \frac{1.2\times10^3}{(1+z)^{-1/4}}\fracb{E_{\rm k,iso,53}}{A_\star}^{1/4}\fracb{\epsilon_{e,-1}}{\xi_e}t_3^{-1/4}
    \quad\quad(k=2) \nonumber
    \label{eq:gamma_fm}
\end{eqnarray}
{at the apparent time $t=(1+z)R/2(4-k)c\Gamma^2=10^3t_3\,$s. Here the factor \linebreak{$\zeta\equiv \Gamma^2ct/R(1+z)$} represents a one-zone 
approximation and is taken here to be $1/2(4-k)$, which is appropriate along the LoS (corresponding to the radial time $t_r$) if 
$\Gamma$ is taken to be that of the shock front, $\Gamma_{\rm sh}$. If~instead, it is taken to be that of the matter just behind 
the shock, $\Gamma\approx\Gamma_{\rm sh}/\sqrt{2}$ then $\zeta=1/4(4-k)$ along the LoS. Since there is significant contribution to 
the observed flux up to angles $\theta\lesssim1/\Gamma$ from the LoS, one should also account for the angular time 
$t_\theta=R/2c\Gamma^2(R)$ along the equal arrival time surface from the shock front. Finally, the~
exact value of $\zeta$ also depends on the effective thickness of the radiating \mbox{shell 
\citep{Waxman-97a,Sari+98,Panaitescu-Meszaros-98a}} and any value is only as good as the one-zone approximation it~represents.}

If the spectral peak (of $\nu F_\nu$) energy of the X-ray flare is $E_x$ in the observer frame, its energy in the comoving 
frame of the blast wave is $E_x'\approx(1+z)\Gamma(1-\beta)E_x=(1+z)E_x/\Gamma(1+\beta)\approx(1+z)E_x/2\Gamma$ for X-ray flare photons 
that are tightly beamed in the radial direction and catch up with the electrons behind the shock with (almost) radial velocity 
vectors. When the forward-shock electrons are in the fast cooling regime, the~peak of the IC spectral component corresponds (without 
accounting for Klein--Nishina effects) to upscattering of $\sim$$E_x$ seed photons (flare photons considered monoenergetic here 
for simplicity) by $\sim\!\gamma_{f,m}$ electrons, e.g., \cite{Wang+06},
\vspace{-6pt}
\begin{eqnarray}
    E_{\rm IC,pk} \approx \gamma_{f,m}^2E_x 
    &\approx& 3.9(1+z)^{3/4}\fracb{\epsilon_{e,-1}}{\xi_e}^2\frac{E_{\rm k,iso,53}^{1/4}E_{x,\rm keV}}{n_0^{1/4}t_3^{3/4}}\,{\rm GeV} 
    \quad\quad(k=0) \\
    &\approx& 1.4(1+z)^{1/2}\fracb{\epsilon_{e,-1}}{\xi_e}^2\frac{E_{\rm k,iso,53}^{1/2}E_{x,\rm keV}}{A_\star^{1/2}t_3^{1/2}}\,{\rm GeV} 
    \quad\quad(k=2)\,. \nonumber
\end{eqnarray}

\textls[+15]{The spectrum of this GeV flash is expected to have  power-law spectral 
indices $d\ln F_\nu/d\ln\nu$ of approximately $-1/2$ and $-p/2$ below and above the energy $E_{\rm IC,pk}$. Klein--Nishina effects 
start to become important for electrons with LF 
$\gamma_e\geq\gamma_{e,\rm KN}= m_ec^2/E_x'=3.3\times10^4(1+z)^{-5/8}E_{\rm k,iso}^{1/8}n_0^{-1/8}t_3^{-3/8}E_{x,\rm keV}^{-1}$ ($k=0$) and 
$\gamma_{e,\rm KN}= 2\times10^4(1+z)^{-3/4}E_{\rm k,iso}^{1/4}A_\star^{-1/4}t_3^{-1/4}E_{x,\rm keV}^{-1}$ ($k=2$), corresponding to $E_{\rm IC}\geq E_{\rm IC,KN}\approx\gamma_{e,{\rm KN}}^2E_x$, and~depending 
on the ratio of electron LFs,}
\vspace{-6pt}
\begin{equation}
    \psi=\frac{\gamma_{e,\rm KN}}{\gamma_{f,m}}=\sqrt{\frac{E_{\rm IC,KN}}{E_{\rm IC,pk}}} \simeq \frac{17}{(1+z)}\fracb{\xi_e}{\epsilon_{e,-1}}E_{x,\rm keV}^{-1}\,,
\end{equation}
the Klein--Nishina suppression of the IC scattered spectrum can occur at energies below or above $E_{\rm IC,pk}$ for $\psi<1$ 
or $\psi>1$, respectively.

\subsection{Pair~Echoes}
\label{sec:Pair-Echoes}
High-energy photons from cosmological sources are absorbed en route by their interaction with the much softer diffuse extragalactic background light (EBL), 
producing $e^\pm$-pairs via $\gamma\gamma$-annihilation \citep{Dai-Lu-02,Razzaque+04b,Wang+04,Murase+07,Ichiki+08,Takahashi+08,Murase+09}. 
The cross section ($\sigma_{\gamma\gamma}$) for the annihilation of two photons \citep{Jauch-Rohrlich-59,Gould-Schreder-67}, with~
energies $E_1$ and $E_2$ colliding with a mutual angle of $\theta_{1,2}$ between their momentum vectors, depends on the (non-dimensional) interaction 
energy $s=E_1E_2(1-\cos\theta_{1,2})/2(m_ec^2)^2$ which must be larger than unity. For~photons traversing through an isotropic radiation bath 
$s\to E_1E_2/(m_ec^2)^2$ for head-on collisions with $\theta_{1,2}=\pi$, 
the~cross section attains its peak value of 
$\sigma_{\gamma\gamma}\approx0.26\sigma_T$ slightly above threshold at $s\approx2$.

VHE primary $\gamma$-ray photons with energy $E$ from GRBs dominantly annihilate with the much softer EBL photons having energy
\vspace{-6pt}
\begin{equation}
    E_{\rm EBL}\approx\frac{2(m_ec^2)^2}{(1+z)^2E}\approx0.5(1+z)^{-2}E_{1\,\rm TeV}^{-1}\,{\rm eV}
\end{equation}
over the mean free path length of
\vspace{-6pt}
\begin{equation}
    \lambda_{\gamma\gamma}(E) = [\sigma_{\gamma\gamma}n_{\rm EBL}(E_{\rm EBL})]^{-1}
    \approx[0.26\sigma_Tn_{\rm EBL}(E_{\rm EBL})]^{-1}\simeq19n_{\rm EBL,-1}^{-1}\,{\rm Mpc}\,.
\end{equation}



The produced $e^-$ and $e^+$ will share the energy of the primary $\gamma$-ray photon equally and have a typical 
LF $\gamma_e=E(1+z)/2m_ec^2\approx10^6(1+z)E_{1\,\rm TeV}$. These pairs will then IC scatter the more numerous and 
softer CMB photons, with~temperature $T_{\rm CMB}(z)=2.73(1+z)\,$K and mean energy 
$E_{\rm CMB}(z)=2.7 k_BT_{\rm CMB}(z)\simeq6.35\times10^{-4}(1+z)$\,eV, to~observed energies
\begin{equation}
    E_{\rm echo} \approx \gamma_e^2 \frac{E_{\rm CMB}(z)}{(1+z)} 
    \simeq 0.6(1+z)^2E_{1\,\rm TeV}^2\,{\rm GeV}\,.
\end{equation}

\begin{figure}[H]
    \centering
   \vspace{-6pt}
 \includegraphics[width=0.45\textwidth]{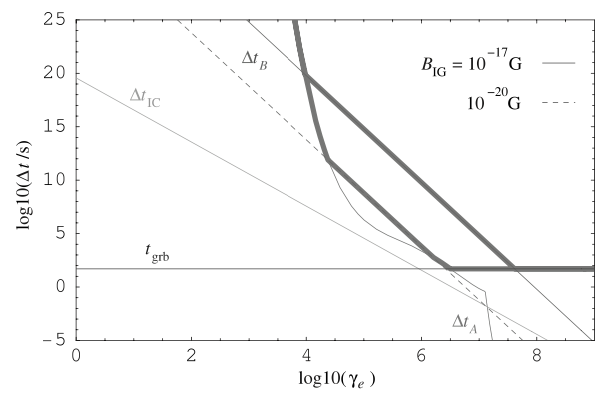} \\
    \includegraphics[width=0.45\textwidth]{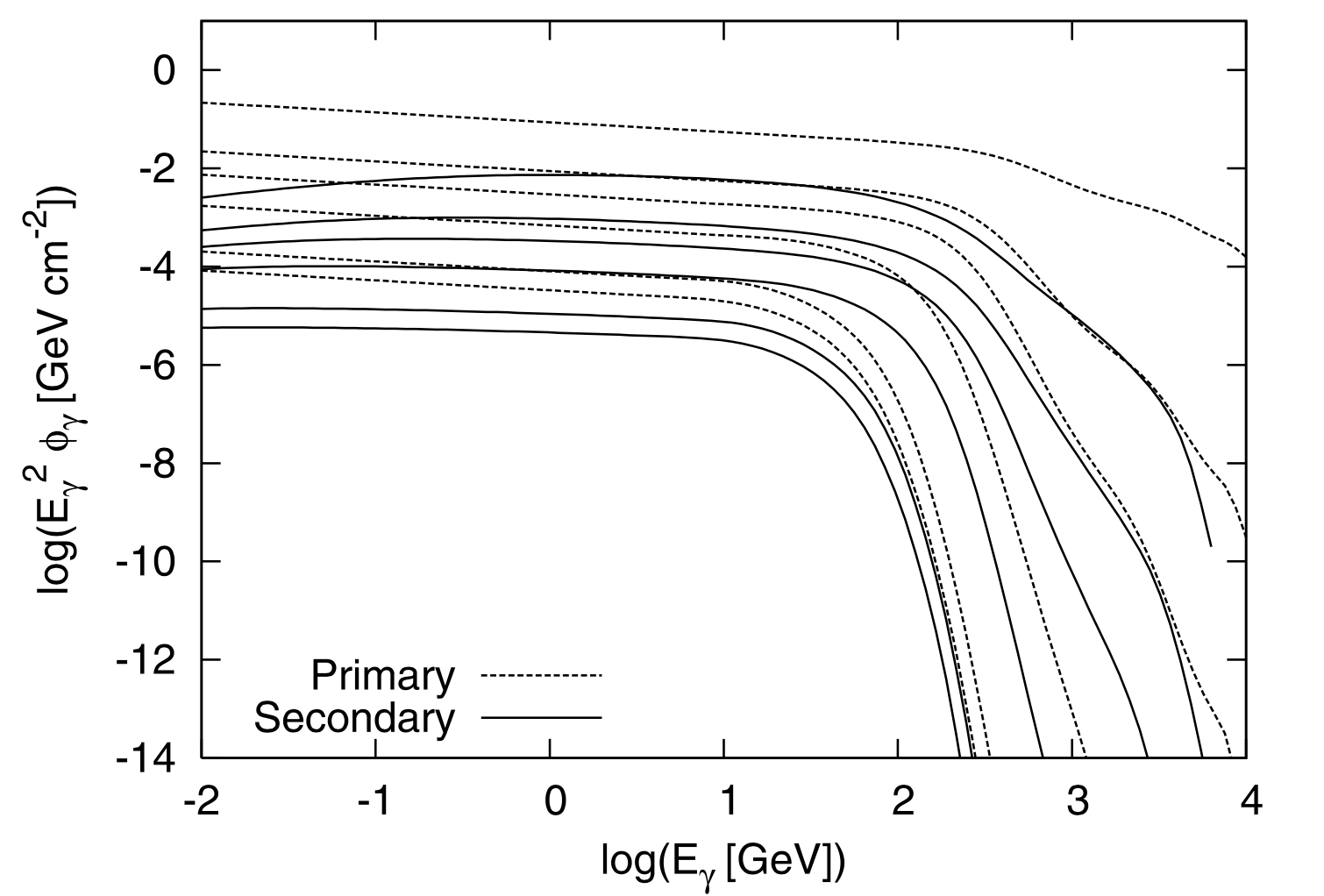} \quad\quad
    \includegraphics[width=0.45\textwidth]{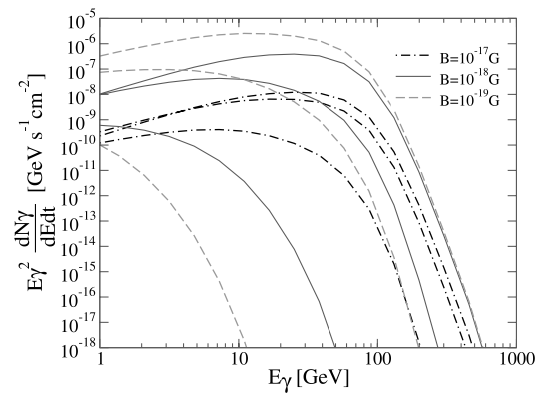}
    \caption{
    {(\textbf{Top})} 
     {Different timescales (thin solid lines) over which the pair echo signal can be temporally 
    smeared, shown for VHE $\gamma$-ray photons produced during the prompt emission over a duration $t_{\rm grb}$. 
    Other timescales are the angular time due to IC cooling ($\Delta t_{\rm IC})$ and deflection of the produced 
    pairs by the IGMF ($\Delta t_B$; shown for two different IGMF magnetic field strengths) and~the angular time 
    associated to the mean free path over which the VHE $\gamma$-ray photons produce pairs ($\Delta t_A$). The~thick 
    solid line highlights the dominant timescale for a given particle Lorentz factor $\gamma_e\simeq1.25\times10^6(E_{\rm echo}/{\rm GeV})^{1/2}\Leftrightarrow E_{\rm echo}\simeq0.64\gamma_{e,6}^2\;$GeV. {Figure from} \citep{Razzaque+04b} (\copyright AAS. Reproduced with permission.).} 
    (\textbf{Bottom-left}) The (observed) model primary and secondary (pair-echo) VHE $\gamma$-ray spectral 
    fluences ($E_\gamma\phi_\gamma=\int F_\gamma dt$, for~
    flux density $F_\gamma$) from GRBs at different redshifts (top to bottom): $z=\{0.1,0.3,0.5,1,3,5\}$. 
    The intrinsic primary spectrum is assumed to be a broken power-law: $dN_\gamma/dE_\gamma\propto(E_\gamma/E_\gamma^b)^{-\alpha}$ for 
    $E_\gamma^{\rm sa}<E_\gamma<E_\gamma^b$ and $dN_\gamma/dE_\gamma\propto(E_\gamma/E_\gamma^b)^{-\beta}$ for 
    $E_\gamma^b<E_\gamma<E_\gamma^{\max}$, where $E_\gamma^{\rm sa}$ is the synchrotron self-absorption break 
    energy, $E_\gamma^b$ is the peak photon energy and $E_\gamma^{\max}$ is the intrinsic high-energy cutoff. 
    The intrinsic spectrum in the figure assumes $\alpha=1$, $\beta=2.2$, $E_\gamma^b=300\,$keV, and~
    $E_\gamma^{\max}=10\,$TeV. {Figure} from \citep{Murase+07} (\copyright AAS. Reproduced with permission.). 
    {(\textbf{Bottom-right}) The observed pair-echo spectrum shown for different IGMF strengths (with coherence length 
    scale $r_{\rm IGMF}=100\,$pc) and at different times ($t_{\rm obs}=10^2\rm~{s},\,10^4\rm~{s},\,10^6\rm~{s}$) 
    for a source at a fixed redshift of $z=1$. The~primary prompt emission spectrum is assumed to be a power-law 
    with photon index $\beta=2.2$ above the peak energy $E_\gamma^b=500\,$keV with a cut-off energy of 
    $E_\gamma^{\max}=10\,$TeV, where the prompt GRB has a duration of $t_{\rm GRB}=50\,$s and luminosity 
    $L_{\gamma,\rm iso}=10^{53}\,{\rm erg\,s}^{-1}$. Figure from \citep{Ichiki+08} (\copyright AAS. Reproduced with permission.).}}
    \label{fig:Pair-Echo-Spectrum}
\end{figure}

This secondary HE emission is dubbed ``Pair Echoes'', and it arrives with a characteristic time delay with respect to the 
primary HE emission due to the pairs being deflected by the weak intergalactic magnetic field (IGMF) {present in cosmic 
voids that are much less dense in comparison to filaments and clusters (much higher and highly structured magnetic fields 
are expected in cosmic filaments ($B\sim10^{-9}$--$10^{-7}$ G \citep{Ryu+98}) and galaxy clusters ($B\sim10^{-7}$--$10^{-6}$ G 
\citep{Kim+91}), where the secondary pairs are expected to produce synchrotron pair echoes \citep{Oikonomou+14} with 
$Y=L_{\rm IC}/L_{\rm syn}\approx U_{\rm CMB}(z)/(B^2/8\pi)\approx10.5(1+z)^4B_{-6}^{-2}$.)}. The~pairs IC cool over a 
characteristic distance
\begin{equation}
    \lambda_{\rm IC,cool}\simeq ct_{\rm IC,cool} 
    = \frac{3m_ec^2}{4\sigma_T\gamma_eU_{\rm CMB}} 
    \approx \frac{0.715\,{\rm Mpc}}{(1+z)^4\gamma_{e,6}}\approx\frac{0.731\,{\rm Mpc}}{(1+z)^5 E_{1\,{\rm TeV}}}\,,
\end{equation}
where $U_{\rm CMB}(z) = aT_{\rm CMB}^4(z)$ is the CMB radiation energy density, and $a$ is the radiation constant. 
Assuming that the pair front expands spherically over a distance $\lambda_{\rm IC,cool}$ with particles at a typical LF $\gamma_e$, 
the radial delay suffered by the secondary HE emission with respect to the primary one is of the order 
$t_{\rm delay} = {(1+z)}\lambda_{\rm IC,cool}(1-\beta_e)/\beta_ec\sim{(1+z)}\lambda_{\rm IC,cool}/2\gamma_e^2c$ for 
$\beta_e\simeq1$ when $\gamma_e\gg1$. The~pair echo will also be temporally smeared out 
but over a much larger angular time $t_{\rm ang} = {(1+z)}(1-\cos\theta)\lambda_{\rm tot}/c\simeq{(1+z)}\theta^2\lambda_{\rm tot}/2c\sim{(1+z)}\lambda_{\rm tot}/2\gamma_e^2c$, 
where $\theta\ll1$ and $\lambda_{\rm tot}=\lambda_{\gamma\gamma}+\lambda_{\rm IC,cool}$, due to light travel time effects over the $\theta\sim1/\gamma_e$ 
angular size of the emission region centered at the observer's line-of-sight. Another angular delay is caused by the deflections of 
the pairs in the intergalactic magnetic field IGMF; ref.~\cite{Plaga-95}. If~the coherence length scale of the IGMF is $r_{\rm IGMF}<\lambda_{\rm IC,cool}$, 
then the root mean square angular deflection is $(\mean{\theta_{B,\rm def}^2})^{1/2} = \kappa_B(\lambda_{\rm IC,cool}/r_{\rm IGMF})^{1/2}(r_{\rm IGMF}/r_L)$, 
{where $\kappa_B$ is an order unity factor that depends on the spectrum of the magnetic field as a function of the coherence length \citep{Ichiki+08,Takahashi+08}}, 
and $r_L = \gamma_em_ec^2/eB_{\rm IGMF}$ is the Larmor radius. The~corresponding angular time over which the pair echo will be smeared is 
$t_{\rm ang, B} \simeq {(1+z)}\mean{\theta_{B,\rm def}^2}\lambda_{\rm tot}/2c$. {For extremely energetic pairs with $\gamma_e\gg1$, the~two timescales,  
$t_{\rm ang}$ and $t_{\rm ang,B}$, can become smaller than $t_{\rm VHE}$, the~duration of the primary VHE emission (which could be either prompt and/or afterglow). 
In addition, for~$\gamma_e\gg1$ 
the mean free path for VHE $\gamma$-ray photons ($\lambda_{\gamma\gamma}$) can become smaller than the cooling distance ($\lambda_{\rm IC,cool}$) of the 
produced pairs, in~which case $\lambda_{\rm tot}\approx\lambda_{\rm IC,cool}$ and $t_{\rm ang}\approx t_{\rm ang,IC}=(1+z)\lambda_{\rm IC,cool}/2\gamma_e^2c$.} 
Therefore, the~correct timescale over which the pair echo signal will be smeared out is 
$t_{\rm ang}+t_{\rm ang,B}+t_{\rm ang,IC}+t_{\rm VHE}\sim\max(t_{\rm ang},t_{\rm ang,B},{t_{\rm ang,IC},t_{\rm VHE}})$. 
{In the top panel of Figure~\ref{fig:Pair-Echo-Spectrum}, we show the different timescales as a function of the particle Lorentz factor $\gamma_e$. 
Only at very large $\gamma_e$ does the timescale $t_{\rm ang,IC}$ dominate $t_{\rm ang}$ due to a sharp decline in $\lambda_{\gamma\gamma}$ caused by the sharp 
rise in the number density of target CMB photons for $\gamma$-rays with $E>$TeV.}

In the bottom-left panel of Figure~\ref{fig:Pair-Echo-Spectrum}, we show example model fluence spectra of the primary and secondary (pair-echo) 
spectra that can be observed from GRBs at 
different redshifts. The~maximum energy of the intrinsic GRB spectrum is assumed to be $E_\gamma^{\max}=10\,$TeV; therefore, the maximum energy of the 
produced pairs is $\sim$5~TeV. As~a result, the~energies of IC scattered CMB photons can reach $\sim$100~GeV, but~these photons may also get absorbed 
en route to us. Above~$\sim$100~GeV, the contribution from IC upscatted CIB photons becomes important, producing an additional bump in the spectrum. 
{The bottom-right panel shows the pair-echo spectrum at different apparent times and for different IGMF strengths from a source at a fixed redshift $z=1$. 
For a given IGMF, the~flux at high energies decays much more rapidly with time compared to the hard power-law at low energies. This is a result of 
shorter IC cooling times and shorter delay times $t_{\rm ang,B}$ for pairs with larger $\gamma_e$. Since weaker IGMFs have shorter $t_{\rm ang,B}$ times, 
the flux is higher initially but decays much faster in comparison to stronger fields that have longer $t_{\rm ang,B}$ times \citep{Ichiki+08}. }

One of the main advantages (see Section~\ref{sec:Non-GRB-Physics} for pair echoes as probes of the IGMF) of detecting pair echoes is that it offers the only 
way to reconstruct the primary VHE emission from GRBs which would otherwise be attenuated due to VHE photons pair producing on EBL~photons.

\subsection{High-Energy $\gamma$-Rays From Pion~Decay}
\label{sec:pion-decay}
Two HE photons are produced directly in the decay of a neutral pion $\pi^0\to2\gamma$, in~which each photon 
escapes with an energy $E_\gamma''=m_{\pi^0}c^2/2\simeq67.5\,$MeV in the rest frame of the pion that is moving 
with LF $\gamma_{\pi^0}$ in the fluid frame. These photons are then detected with energy 
$E_\gamma(1+z) \sim \Gamma\gamma_{\pi^0}E_\gamma''\gtrsim7\Gamma_2\,$GeV. Neutral pions can be produced via the 
following collisional processes between protons ($p$), neutrons ($n$), and~photons 
($\gamma$):
\begin{equation}
    p+p \to p+p+\pi^0, \quad p+n\to p+n+\pi^0, \quad p+\gamma\to \Delta^+\to\pi^0+p
\end{equation}

The most important of the above hadronic energy-loss mechanisms is the photohadronic process, where a 
photon interacts with a proton, at~a threshold photon energy of 
$E_{\gamma,\rm th}''=(m_\pi+m_\pi^2/2m_p)c^2\simeq150\,$MeV in the rest frame of the proton, to~produce a pion, 
\mbox{e.g., \cite{Dermer-Atoyan-06}.} When a typical $\gamma$-ray photon with energy 
$E_\gamma'=(1+z)E_\gamma/\Gamma$ interacts with a proton in the flow having LF $\gamma_p$, the~scattering 
cross section for the $\Delta^+$ resonance peaks when the energy of the photon 
in the proton's rest frame is $E_\gamma''=E_\gamma'\gamma_p(1-\beta_p\mu_{p\gamma})\simeq0.3\,$ GeV~\citep{Mucke+99}, where $\beta_p=(1-\gamma_p^{-2})^{1/2}$ and $\mu_{p\gamma}=\cos\theta_{p\gamma}$. This is equivalent to the proton 
having energy 
$E_p'=\Gamma E_\gamma''m_pc^2/(1+z)E_\gamma(1-\beta_p\mu_{p\gamma})\sim \Gamma(0.3\,{\rm GeV}^2)/E_\gamma(1+z)$. 
If $E_{\pi^0}'\sim0.2E_p'$, it would yield a VHE $\gamma$-ray photon of energy
\begin{equation}
    E_{\gamma,\rm VHE} = \frac{\Gamma}{(1+z)}\frac{E_{\pi^0}'}{2}\sim\frac{300}{(1+z)^2}\Gamma_2^2\fracb{E_\gamma}{1\,{\rm MeV}}^{-1}\,{\rm TeV}\,.
\end{equation}

The above three collisional processes also produce charged pions ($\pi^+$ and $\pi^-$), which then decay to muons 
that further decay to produce electrons and positrons that can then produce HE synchrotron photons. The~most 
important for producing HE photons is again the $\Delta^+$ resonance that also yields
\begin{equation}
    \Delta^+\to\pi^++n,\quad \pi^+\to\mu^++\nu_\mu\to e^++\nu_e+\bar\nu_\mu+\nu_\mu\,,
\end{equation}
where $\mu^+$ is the anti-muon and $\nu_\mu$ and $\bar\nu_\mu$ are its neutrino and anti-neutrino, and~$\nu_e$ is the electron neutrino. Approximately 20\% of 
the proton's energy goes into $\pi^+$, which is further equally distributed between the pion's decay products \citep{Waxman-Bahcall-97}. This produces 
a  high-energy positron with LF $\gamma_+\sim0.05E_p'/m_ec^2 \simeq 3\times10^6(1+z)^{-1}\Gamma_2(E_\gamma/1\,{\rm MeV})^{-1}$ that produces HE synchrotron 
photons of energy
\begin{equation}
    E_{\gamma,\rm VHE} = \frac{\Gamma}{(1+z)}\gamma_+^2\frac{\hbar eB'}{m_ec} 
    \simeq \frac{1}{(1+z)^3}\fracb{E_\gamma}{1\,{\rm MeV}}^{-2}B_5'\Gamma_2^3\,{\rm TeV}\,.
\end{equation}

The photo-hadronic process, if~operating in GRBs, opens up prospects for detecting high-energy ($\sim10^{14}$~eV) neutrinos by km-scale 
ground-based detectors \citep{Waxman-Bahcall-97}, e.g.,~IceCube~\citep{Ahrens+04}. The~intrinsic ratio between muon and electron neutrinos 
{at the source is expected to be} 2:1 (with no $\tau$ neutrinos), {but vacuum oscillations between the three neutrino flavors 
($\nu_e,\nu_\mu,\nu_\tau$) may yield equal distributions at Earth. The~intrinsic ratio at the source can be different from 1:2:0 when the 
neutrinos are produced inside a star, e.g.,~a jet that propagates inside a blue supergiant. In~this case, resonant flavor oscillations 
in matter due to the Mikheyev--Smirnov--Wolfenstein effect will alter the intrinsic ratios at the source \citep{Mena+07}; therefore, after 
vacuum oscillations, the ratios observed at Earth will also be different from 1:1:1. The~intrinsic flavor ratios can further be modified at 
high energies due to electromagnetic and adiabatic energy losses of muons and pions \citep{Kashti-Waxman-05} as well as due to matter \mbox{oscillations 
\citep{Mena+07,Razzaque-Smirnov-10}} at the source. This would lead to energy-dependent, unequal flavor ratios measured at Earth.} 

Detection of neutrinos from GRBs can only happen for very bright GRBs with 
$\gamma$-ray fluences $\gtrsim10^{-4}\,{\rm erg\,cm}^{-2}$ \citep{Dermer-Atoyan-03,Razzaque+04a,Asano-05}. Thus, far neutrino searches by IceCube have come out 
empty, even in the case of very bright GRBs, e.g.,~130427A \citep{Gao+13}, and~deeper upper limits have offered strong constraints on GRB physics and 
neutrino production therein \citep{Aartsen+17,Albert+21}.

Apart from the $e^\pm$-pairs produced in the photo-hadronic cascades, additional secondary $e^\pm$-pairs result from $\gamma\gamma\to e^+ + e^-$ that 
can have important effects on both the low and high-energy parts of the spectrum \citep{Asano+09a,Asano+09b}. Such pair cascades can also be important in other 
hadronic scenarios, namely proton synchrotron emission as discussed in Section~\ref{sec:proton-synchro}. The~injected pair spectrum in this case has 
$d\ln n_e/d\ln\gamma_e\simeq-2$ which yields a relatively flat $\nu F_\nu$ synchrotron spectrum. At~low energies, i.e.,~below the peak of the sub-MeV 
Band component, the~synchrotron emission from secondary pairs might dominate and make the spectrum too soft when compared with observations (see 
bottom-left panel of {Figure}~4). {However, if~the secondary pairs are stochastically accelerated (or heated) by MHD/plasma turbulence, 
then a low-energy photon index of $\alpha\sim-1$ that matches observations can be produced \citep{Murase+12}.} Above the Band-component peak energy, 
the spectrum is modified due to IC scattered emission by the secondary~pairs.

\subsection{{High-Energy $\gamma$-Rays from the Bethe--Heitler~Process}}

{
The Bethe--Heitler process is a photo-hadronic interaction in which the $e^\pm$-pairs are produced directly,
\begin{equation}
    p+\gamma\to p + e^- + e^+\,.
\end{equation}

The differential cross section for this process \citep{Bethe-Maximon-54,Chodorowski+92} strongly depends on the angle $\theta_\pm'$ 
between the incoming photon and the outgoing $e^\pm$ in the proton's rest frame. It peaks sharply near $\theta_\pm'\sim1/\gamma_\pm'$, 
where $\gamma_-'$ ($\gamma_+')$ is the LF of the electron (positron) in the proton's rest frame. When the proton's LF in the jet comoving 
frame, $\gamma_p$, is much larger than that of the produced pairs in the proton's rest frame, with~$\gamma_p\gg\gamma_\pm'\gg1$, then 
the pairs are produced with LF in the jet comoving frame of  $\gamma_\pm=\gamma_p\gamma_\pm'(1-\beta_p\beta_\pm'\cos\theta_\pm')\approx(\gamma_p\gamma_\pm'/2)(\gamma_p^{-2}+\gamma_\pm'^{-2}+\theta_\pm'^2)\approx\gamma_p/\gamma_\pm'$, where on average $\gamma_\pm\sim\gamma_p/5$ \citep{Crumley-Kumar-13}. For~typical prompt emission spectral peak energies of 
$E_{\rm pk}\lesssim m_ec^2/(1+z)$, the~Bethe--Heitler process is less efficient (by a factor of $\sim10^2$) in producing pairs compared 
to the $\Delta^+$ resonance when the LF of produced pairs is $\gamma_\pm\gtrsim10^6$. However, for~$\gamma_\pm\lesssim10^3$, it can be 
much more efficient, while for $10^3\lesssim\gamma_e\lesssim10^6$ its efficiency depends on the spectral index of the prompt emission 
\citep{Crumley-Kumar-13}. The~high-energy pairs produced in the process can then give rise to HE to VHE photons via synchrotron or IC emission.
}

\section{GRB Prompt HE Emission---Observations vs.~Theory}
\label{sec:Prompt-HE-Emission}

\begin{figure}[H]
    \includegraphics[width=0.7\textwidth]{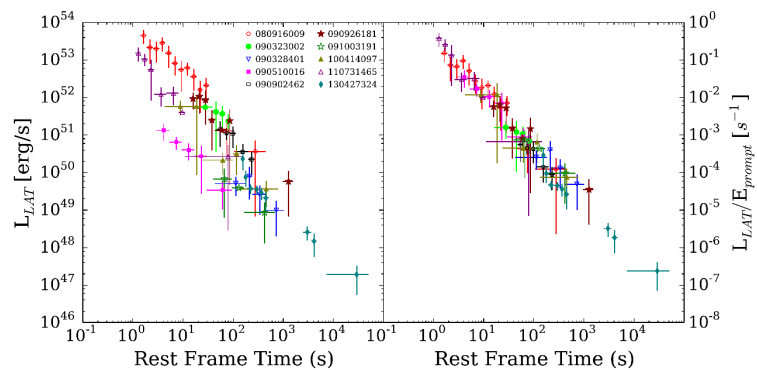}
    \caption{{(\textbf{Left}):}   \textit{Fermi}-LAT GRB lightcurves; (\textbf{Right}): The same as the left-panel but normalized by the energy released during 
    the prompt emission. {Figure from} \citet{Nava+14} (also {see} \citep{Ghisellini+10}).}
    \label{fig:LAT-GRBs-Lightcurves}
\end{figure}

HE emission in the energy range of (0.1--100)~GeV has been detected by the \textit{Fermi}-LAT in more than 170 GRBs \citep{Ajello+19}. 
Prior to \textit{Fermi}, emission in this energy range (but below $\sim$20~GeV) was also detected by the Energetic Gamma Ray Experiment 
Telescope (EGRET) aboard the now defunct Compton Gamma-Ray Observatory (CGRO) from a handful of GRBs \citep{Schneid+92,Hurley+94,Dingus-01,Gonzalez-03}. 
In most of the \textit{Fermi}-LAT 
GRBs and that detected by EGRET, the~broadband prompt emission spectrum is described by a single Band-like spectral component, generally peaking 
in the (0.1--1)~MeV range and also extending to high energies. In~rare cases, it shows a clear spectral cutoff in the (20--350)~MeV energy range 
that is interpreted as a result of the opacity of HE photons to $\gamma\gamma$-annihilation within the source \citep{Tang+15,Vianello+18}. This may 
be intrinsically more common, the~low observed fraction being a matter of sensitivity, as~such a cutoff appears in 20\% out of a sample of bright 
\textit{Fermi} Gamma-ray Burst Monitor (GBM) bursts when performing a joint GBM-LAT spectral fit \citep{Ackermann+12}. Moreover, many 
of the bright \text{Fermi}-LAT GRBs show a second spectral component, in~addition to the softer Band-like component, that dominates the HE 
emission \citep{Abdo+09b,Ackermann+10,Ackermann+11}, typically well fit by a power law and sometimes showing a cutoff (also likely due to intrinsic 
$\gamma\gamma$-opacity). A~similar HE spectral component was also seen by EGRET \citep{Gonzalez-03}. Overall, the~HE emission 
seen by the \textit{Fermi}-LAT shows three remarkable features \citep{Ajello+19}: 
\begin{enumerate}
    \item \textbf{Extra HE spectral component}: An extra power-law spectral component that extends to high energies and which is distinct from the 
    typical sub-MeV Band component appears in several bright LAT detected GRBs.
    \item \textbf{Delayed onset}: The onset of this HE emission is delayed relative to the softer $\gamma$-rays near the spectral peak, with~typical delays of a few to 
    several seconds ($0.1\,{\rm del}\lesssim t_{\rm del}\lesssim40\,$s) for long-soft GRBs and~a few tenths of a second ($0.05\,{\rm s}\lesssim t_{\rm LAT}\lesssim1\,$s) 
    for short-hard GRBs. 
    \item \textbf{Long duration}: It is systematically longer-lasting ($t_{\rm LAT}\lesssim35\,$ks) with its flux decaying smoothly as a power law 
    in time (see Figure~\ref{fig:LAT-GRBs-Lightcurves}) having $d\log F_\nu/d\log t\sim-1$ \citep{Ackermann+13}.
\end{enumerate}
\vspace{-9pt}

In the following, we discuss possible origins of the HE spectral component and its delayed onset with respect to the sub-MeV emission 
(also see, e.g.,~\citep{Gehrels-Razzaque-13} for a review).

There are two main emission regions from where the HE spectral component can be produced. The~first is internal to the outflow in which the emission 
arises due to dissipation of kinetic energy, e.g.,~via internal shocks, or~magnetic energy, e.g.,~due to magnetic reconnection, and~it occurs at smaller 
radii before the outflow is significantly slowed down by its interaction with the circumburst medium. In~this case, the~emission is expected to be 
highly variable, with~$t_v/t\ll1$ where $t_v$ is the variability timescale, and~correlated with the sub-MeV prompt emission, which is 
seen in all cases (as in this case, the two arise from the same outflow, albeit possibly at different radii). The~second region is the external forward 
(afterglow) shock in which case the emission is produced by shock-heated swept-up circumburst medium. 
In contrast to the prompt emission, the~lightcurve is expected to be much smoother, with~$t_v\sim t$, and~decaying after its peaks at 
$t\gtrsim t_{\rm GRB}$. Such behavior was also observed in many cases. In~many LAT GRBs, there is initially a variable GeV emission followed by a 
smooth tail with a spectral change in the transition, suggesting a transition between prompt and afterglow GeV emission. Furthermore, upon~closer inspection, 
in many cases the delayed onset is caused by the fact that the first spike in the prompt GRB lightcurve is missing at $\sim\,$GeV energies, and~only 
subsequent spikes appear in $\sim\,$GeV and coincide with those at sub-MeV~energies.

\subsection{Delayed Onset of the \textit{Fermi}-LAT HE~Emission}
\label{sec:Delayed-Onset}

In both the long-soft and short-hard GRBs detected by \textit{Fermi}-LAT, the~HE emission is generally delayed by $t_{\rm del}\sim$~(0.1--40)~s in the former 
and $t_{\rm del}\sim$~(0.05--1)~s in the latter. While formally $t_{\rm del}$ reaches values as high as $\lesssim$$10^4$~s in rare cases of both populations 
\citep{Ajello+19}, these are mostly cases where the GRB was outside the LAT FoV at the time of the GRB trigger and~likely do not have a similar physical origin. 
In the majority of GRBs, the~onset of LAT HE emission occurs before the softer prompt $\gamma$-ray emission recorded by \textit{Fermi}-GBM is over. A~number of 
different scenarios have been proposed to explain the delayed onset, which we briefly discuss~below. 

\subsubsection{Forward External Shock~Emission} 
The shock-heated electrons behind the forward shock radiate synchrotron photons that produce the broadband afterglow emission, whose lightcurve peaks 
at the apparent time (assuming a thin-shell case, for~which $t_{\rm dec}>t_{\rm GRB}$)
\begin{equation}
    t_{\rm dec} = (1+z)\frac{R_{\rm dec}}{2c\Gamma_0^2} = \begin{cases}
    18\fracb{1+z}{2}E_{53}^{1/3}n_0^{-1/3}\Gamma_{0,2.5}^{-8/3}\,{\rm s} & (k=0) \\
    5.9\fracb{1+z}{2}E_{53}A_\star^{-1}\Gamma_{0,2}^{-4}\,{\rm s} & (k=2)\,.
    \end{cases}
\end{equation}
   
In this scenario \citep{Zou+09,Kumar-Barniol-Duran-09,Kumar-Barniol-Duran-10,Ghisellini+10,Granot-12a}, $t_{\rm dec}$ 
is the relevant timescale to explain $t_{\rm del}$. Furthermore, for~$R>R_{\rm dec}$, the~proper velocity of the blast 
wave starts to decline as $u(R)=\Gamma(R)\beta(R)\propto R^{-(3-k)/2}$ as more mass is swept up, and the dynamical 
evolution of the blast wave becomes self-similar \citep{Blandford-McKee-76}. For~an \textit{adiabatic} (constant energy 
with negligible radiative losses) relativistic spherical blast wave, the flux density for $\nu>\max(\nu_m,\nu_c)$, the~
frequency regime relevant for HE afterglow emission, scales as $F_\nu(t)\propto\nu^{-p/2}t^{-(3p-2)/4}$ for $t>t_{\rm dec}$~\citep{Sari+98,Granot-Sari-02}. If~the blast wave is \textit{radiative} (a short-lived early phase where its energy 
decreases over time due to radiative losses), the flux density has the scaling $F_\nu(t)\propto\nu^{-p/2}t^{-(6p-2)/7}$ 
\citep{Sari+98}.  (Even for fast cooling only a fraction $\epsilon_e$ of the 
internal energy generated at the afterglow shock is radiated away, and~a similar fraction of the total energy is radiated over each dynamical time, so 
the blast wave may be far from being fully radiative as assumed in this scaling.) Evidently, the~forward shock emission 
generally obeys a closure relation, whereby the temporal and spectral indices are coupled by virtue of their dependence on the electron energy distribution 
power-law index $p$. \citet{Kumar-Barniol-Duran-09,Kumar-Barniol-Duran-10} showed that three \textit{Fermi}-LAT GRBs obeyed this closure relation of an adiabatic 
blast wave, with~$p=2.4\pm0.06$ (GRB 080916C) and $p=2.2\pm0.2$ (GRB 090510, 090902B), that yielded $d\log F_\nu/d\log t = -(3p-2)/4 = 1.15-1.3$ consistent 
with the observed value to within 1-$\sigma$ uncertainty. An additional argument in favor of this scenario is that the LAT emission lightcurve shows a very smooth 
decay, which is expected for afterglow emission. A~caveat here is that this applies mainly to the long-lived LAT emission at $t\gtrsim t_{\rm GRB}$, whereas the variable $\sim\,$GeV emission seen at $t\leq t_{\rm GRB}$ in bright LAT GRBs cannot be afterglow emission and is most likely prompt emission (especially when it is temporally correlated with $\lesssim\,$MeV spikes in the prompt GRB lightcurve, e.g.,~\citep{Abdo+09a,Abdo+09b,Abdo+09c,Ackermann+10,Ackermann+11})

\begin{figure}[H]
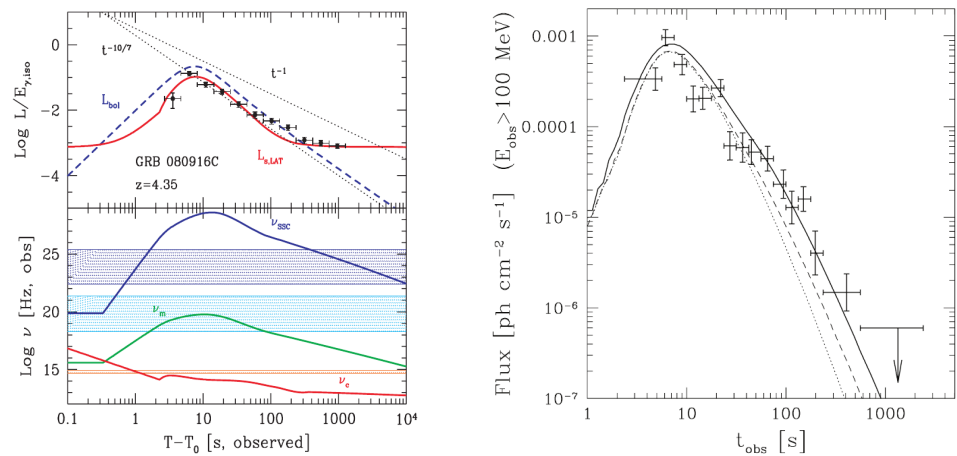

    \includegraphics[width=0.39\textwidth]{figs/080916C-Ghisellini+10.pdf}\quad\quad
    \includegraphics[width=0.45\textwidth]{figs/080916C-Beloborodov+14.pdf}
    \caption{(\textbf{Left}) Radiative afterglow with $e^\pm$ enrichment model fit to the lightcurve of GRB {{080916C}}~\citep{Ghisellini+10}. In~the top-panel, the~
    red curve shows the \textit{Fermi}-LAT luminosity, the~dashed blue curve is the expected bolometric luminosity, and~the two dotted black 
    curves show the expected temporal slopes of the afterglow luminosity when the blast wave is adiabatic ($L_{\gamma,\rm iso}\propto t^{-1}$) or 
    radiative ($L_{\gamma,\rm iso}\propto t^{-10/7}$) for a uniform circumburst medium. The~bottom panel shows the temporal evolution of the 
    characteristic frequencies of synchrotron and SSC emission. The~shaded regions show the energy ranges of the LAT [($0.1-100$)\,GeV] and~
    GBM [($8-10^3$)\,keV] instruments, as~well as the optical energy range (U and R filters). 
    \textbf{Right}: Pre-accelerated and pair-loaded CBM afterglow model fit to the \textit{Fermi}-LAT lightcurve of GRB 080916C \citep{Beloborodov+14} (\copyright AAS. Reprodued with permission.).}
    \label{fig:080916C-LC-Model-Fit}
\end{figure}

The temporal evolution of the observed (isotropic-equivalent) luminosity is another useful probe for the origin of the HE emission. From~energy 
conservation, $E\propto\Gamma^2R^{(3-k)}$, and~with $\Gamma\propto R^{(k-3)/(1+\delta)}$ where $R\propto\Gamma^2t$, the~time evolution of the 
isotropic-equivalent energy can be obtained, $E_{\rm iso}\propto t^{[(\delta-1)(3-k)]/(7+\delta-2k)}$ \citep{Meszaros+98}. For~an adiabatic ($\delta=1$) 
blast wave, $E_{\rm iso}\propto t^0$, as~it should be, and~for radiative ($\delta=0$) blast wave $E_{\rm iso}\propto t^{(k-3)/(7-2k)}$. The~observed 
luminosity then follows with $L_{\gamma,\rm iso}\propto E_{\rm iso}/t \propto t^{[(\delta-3)(3-k)-(\delta+1)]/(7+\delta-2k)}$, which yields 
$L_{\gamma,\rm iso}\propto t^{-1}$ for the adiabatic case and $L_{\gamma,\rm iso}\propto t^{(3k-10)/(7-2k)}$ for the radiative case. The~left panel 
of Figure~\ref{fig:080916C-LC-Model-Fit} shows the radiative afterglow model fit to the LAT lightcurve of GRB 080916C \citep{Ghisellini+10}. This agreement 
presents a strong argument in favor of the synchrotron afterglow origin of the late-time LAT HE emission. However, the~main LAT peak is too sharp 
to arise from the afterglow onset corresponding to the outflow deceleration time, and~instead matches the second $\sim\,$MeV peak, so it is more 
likely associated with the prompt GRB emission, while the temporally smoother afterglow GeV emission likely starts dominating later, after~a few tens 
of seconds, with~a rather shallow decay slope ($\sim t^{-1}$).

There are two major hurdles for this scenario. First, many LAT GRBs show a peak in the GeV emission while the prompt emission is still active, which 
is difficult to explain with synchrotron emisson from the external forward shock. In~the thin-shell afterglow shock scenario \citep{Sari-Piran-95}, 
the peak of the HE emission will occur at $t = t_{\rm dec} = (1+z)R_{\rm dec}/2c\Gamma_0^2$ which is always larger than the duration of the prompt GRB 
emission, $T_{\rm GRB}=(1+z)\Delta_0/c$, given by the shell crossing time of the ejecta shell of initial thickness $\Delta_0$. Alternatively, in~the 
thick-shell case, $t_{\rm dec}\sim T_{\rm GRB}$. Second, this model cannot explain the detection of VHE photons at late times when $t > t_{\rm dec}$ where 
the detected photons have energies much larger than $E_{\rm syn,max}$ \citep{Ackermann+14}. Both of these arguments suggest that yet another mechanism 
might be responsible for the LAT~emission.

\subsubsection{Inverse-Compton GeV~Flash}
The shock-heated electrons behind the forward shock at radius $R$ can be Compton-cooled by prompt emission $\sim\,$MeV photons emitted at a smaller 
radius $R_{\rm prompt}\ll R$ as the radiation front overlaps with the blast wave \citep{Beloborodov-05a,Beloborodov-05b,Beloborodov+14}. When the 
prompt emission photons travel ahead of the blast wave, a~small fraction is scattered by the 
yet unshocked electrons in the CBM at large angles from the radial direction. The~scattered photons then produce $e^\pm$-pairs via $\gamma\gamma$-annihilaton 
on the radially expanding (collimated) prompt emission radiation front. The~created pairs further scatter the prompt photons, causing exponential pair-creation 
and the resultant high multiplicity (with $\mathcal{M}_\pm\lesssim10^5$) pair-loading of the CBM ahead of the forward shock 
\citep{Thompson-Madau-00,Meszaros+01,Beloborodov-02,Kumar-Panaitescu-04,Thompson-06}. Scattering of the prompt radiation 
by the pair-loaded CBM also imparts momentum to the pairs and pre-accelerates them to a typical LF 
$\gamma_{\rm pre} = (1-\beta_{\rm pre}^2)^{-1/2} < \Gamma_{\rm bw}$, where $\Gamma_{\rm bw} = (1-\beta_{\rm bw}^2)^{-1/2}$ is the bulk LF of the blast wave, 
i.e., material just behind the forward shock. As~the blast wave sweeps up the pair-loaded CBM, with~a relative LF 
$\Gamma_{\rm rel}=\Gamma_{\rm bw}\gamma_{\rm pre}(1-\beta_{\rm bw}\beta_{\rm pre})\approx\Gamma_{\rm bw}/\gamma_{\rm pre}(1+\beta_{\rm pre})$, 
the shock-heated pairs are thermalized with $\gamma_{\rm th}\sim\Gamma_{\rm rel}$ when $\mathcal{M}_\pm\gg10^3$. This model assumes that only 
a small number of particles are accelerated into a power-law energy distribution and most of the energy resides with the quasi-thermal pairs. 
The radiative efficiency of the shock-heated pairs is almost 100\% during the GeV flash; therefore, the blast wave does not start to evolve 
adiabatically until all the prompt emission photons have overtaken~it. 

In this scenario, the~peak of the LAT emission occurs at $t_{\rm pk}=(1+z)R_{\rm pk}/2c\Gamma_{\rm pk}^2$ where $R_{\rm pk}\sim10^{16}\,$cm (in the case of 
GRB 080916C) is the radius where the LF of the electrons behind the forward shock is
\begin{equation}
    \gamma_{\rm th,pk}\sim50\fracb{E_{\rm IC}}{1\,{\rm GeV}}^{1/2}\fracb{E_{\rm prompt}}{1\,{\rm MeV}}^{-1/2}\,,
\end{equation}
so that the IC scattered emission peaks in the GeV energy range. At~$R<R_{\rm pk}$, the contrast between $\Gamma_{\rm bw}$ and $\gamma_{\rm pre}$ is small and 
therefore $\gamma_{\rm th}<\gamma_{\rm th,pk}$. This contrast grows over larger radii and $\gamma_{\rm th}=\gamma_{\rm th,pk}$ at $R=R_{\rm pk}$, and~for 
$R>R_{\rm pk}$, the contrast is much larger which yields $\gamma_{\rm th}>\gamma_{\rm th,pk}$ and produces VHE emission at $\gtrsim$TeV energies. The~right panel 
of Figure~\ref{fig:080916C-LC-Model-Fit} shows the model fit to the LAT lightcurve of 080916C from \citet{Beloborodov+14}.

\subsubsection{Synchrotron Emission from Protons Accelerated at the External Forward~Shock}
In the hadronic scenario, a~proton-synchrotron emission model with a strong comoving B-field ($B'$) can explain the delayed 
onset of the LAT emission \citep{Razzaque+10,Razzaque-10,Asano-Meszaros-12}. Just like electrons, protons are also accelerated at the 
external blast wave to energies where they can radiate $\sim\,$GeV to TeV synchrotron radiation. This radiation is further 
processed into $e^\pm$-pairs via $\gamma\gamma$-annihilation where the produced pairs then radiate sub-GeV 
synchrotron photons. The~onset of HE emission is delayed due to two effects. First, protons are accelerated over the Larmor 
time to achieve a maximum LF $\gamma_{M,p}$ (see Section~\ref{sec:proton-synchro} for definition), which causes a delay of at least
\begin{equation}
    t_{\rm L,p} \approx \frac{(1+z)}{\Gamma}\frac{\gamma_{M,p}m_pc}{eB'} 
    = \frac{(1+z)}{\Gamma}\frac{m_p^2c}{m_e}\fracb{6\pi}{e\sigma_TB'^3}^{1/2}
    \approx \frac{0.2(1+z)}{B_4'^{3/2}\Gamma_2}\,{\rm s}\,.
\end{equation}

This should also be a lower limit on the variability time, as~the local emission cannot turn on or off faster than this.
Second, as~shown in the model put forth by \citet{Razzaque+10}, it takes a finite amount of time for the peak of the 
proton synchrotron radiation spectrum, which peaks at higher energies at early times, to~move into the LAT energy~range. 

This scenario requires a strong magnetization of the shocked material downstream of the blast wave to explain the delays in the LAT emission 
onset. 
A major weakness of this model is that it is radiatively inefficient and~therefore requires a large amount of energy in accelerated 
protons that must be injected with minimum LF of $\gamma_{m,p}\gtrsim 10^6$~\citep{Wang+09,Crumley-Kumar-13}. Furthermore, as~the 
proton synchrotron cooling break sweeps across the observed energy band, the~spectral index should change from $d\ln F_\nu/d\ln\nu = (1-p)/2$ to 
$-p/2$, where $p$ is the power-law index of the proton energy distribution, $n_p(\gamma_p)\propto\gamma_p^{-p}$ for 
$\gamma_{p,\rm min}\leq\gamma_p\leq\gamma_{p,\rm max}$. However, no such spectral change has been observed in the delayed LAT emission. 
Although it is possible that this spectral component has only been observed at energies above the cooling break due to the limiting sensitivity 
of the \textit{Fermi}-LAT at high energies, it would be too much of a coincidence to have happened in all LAT bursts that show delayed~emission.

\subsubsection{SSC~Emission}
The delay time of the LAT emission in the SSC scenario depends on the time it takes for the IC-scattered radiation field to 
build up in the LAT energy band. That depends on the temporal evolution of the Compton-$y$ parameter which must become larger 
than unity for IC scattering to become the dominant particle cooling mechanism. Detailed one-zone numerical simulations \citep{Bosnjak+09,Asano-Meszaros-11,Asano-Meszaros-12} of prompt GRB emission show that under certain conditions SSC emission 
in the LAT energy range can be delayed with respect to the sub-MeV synchrotron component due to the time it takes to build up 
the seed synchrotron photon field in the emitting region (of the order of its light crossing time). However, in~many cases, the 
temporal delay is insufficient to explain the observed ones and remains limited to $t_{\rm delay} < t_v$, where $t_v$ is the 
variability timescale. This effect would also lead to a systematic delay of the GeV emission w.r.t the sub-MeV emission 
for each spike in the prompt lightcurve. In~practice, the~observed delay typically reflects the first spike being absent in 
the GeV, with~subsequent spikes coinciding in MeV and~GeV.



\subsection{Distinct HE Spectral~Component}

\begin{figure}[H]
    \includegraphics[width=0.45\textwidth]{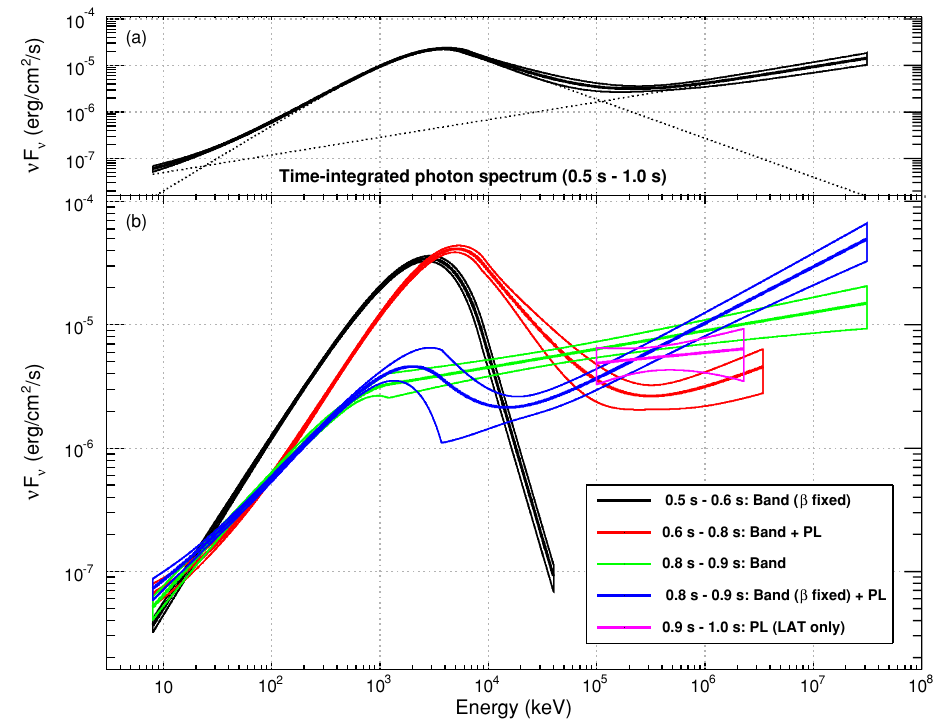}\quad\quad
    \includegraphics[width=0.45\textwidth]{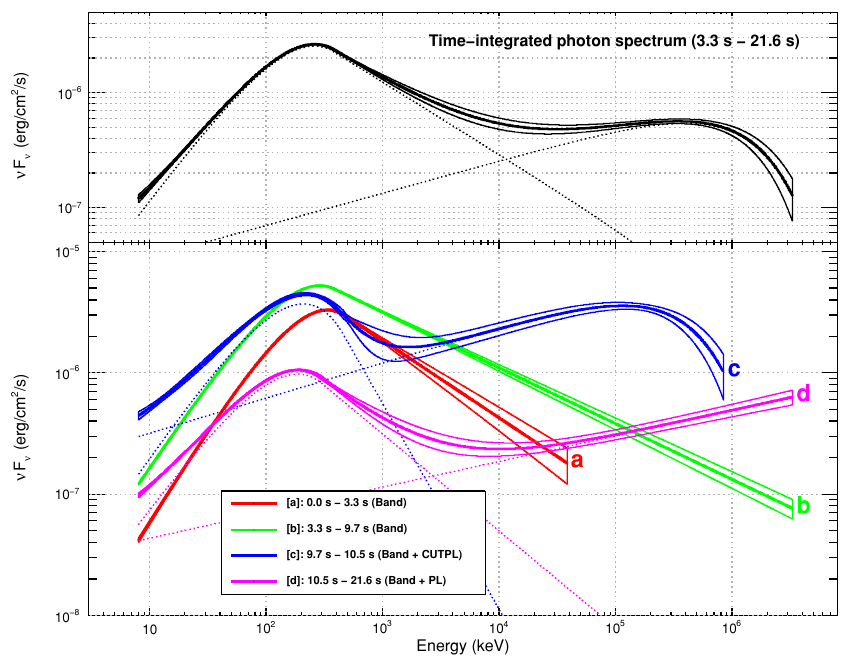}
    \includegraphics[width=0.45\textwidth]{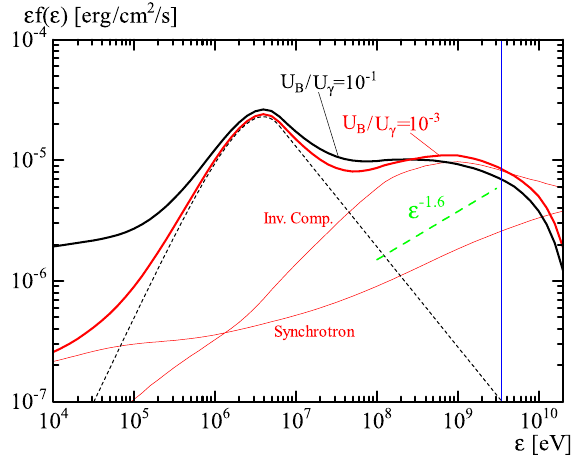}\quad\quad
    \includegraphics[width=0.45\textwidth]{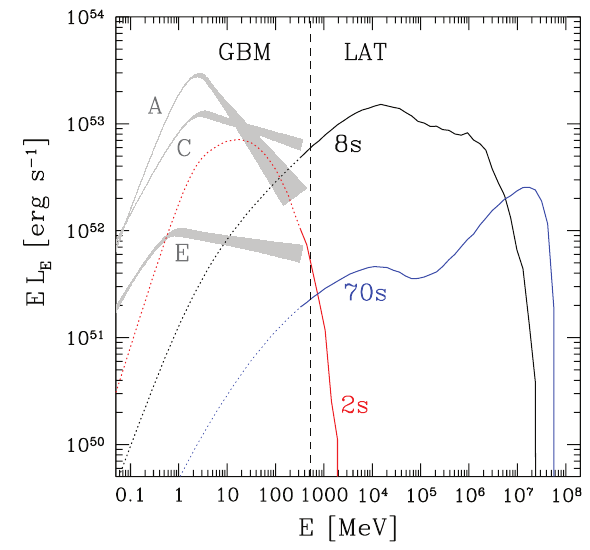}
    \caption{\textbf{Top}: Time-integrated and time-resolved prompt emission spectra of short-hard GRB 090510 ({left;} \citep{Ackermann+10}, \copyright AAS. Reproduced with permission.) 
    and long-soft GRB 090926A ({right;} \citep{Ackermann+11}, \copyright AAS. Reproduced with permission.). In~both cases, the~spectrum shows a low- and high-energy excess 
    which is fit by a power-law or cutoff power-law component that has a distinct origin from the main Band component. 
    \textbf{Bottom}: Theoretical modeling of the spectrum of GRB 090510 using a hadronic scenario with photo-hadronic cascades ({left;} \citep{Asano+09b}, \copyright AAS. Reproduced with permission) and 
    that of GRB 080916C using the pre-accelerated and pair-loaded ISM in the afterglow model ({right;} \citep{Beloborodov+14}, \copyright AAS. Reproduced with permission.).}
    \label{fig:PL-Component}
\end{figure}

Many bright \textit{Fermi}-LAT GRBs show a distinct HE spectral component in addition to the Band-like spectrum, where the 
latter represents the canonical prompt emission spectrum peaking in the $\sim$(0.1--1)~MeV energy range. This additional component has been modeled as a power-law, sometimes with a high-energy cutoff, in~addition to the Band component.
Such a component was required by the data 
in GRB 090227B \citep{Guiriec+10}, GRB 090228~\citep{Guiriec+10}, \mbox{GRB 090510 \citep{Ackermann+10}}, GRB 090902B \citep{Abdo+09b}, GRB 090926A \citep{Ackermann+11}, 
GRB \mbox{110731A \citep{Ackermann+13}}, \linebreak GRB 130427A \citep{Ackermann+14}, GRB 141207A \citep{Arimoto+16}, GRB 190114C \citep{Ajello+20}. The~first such detection of 
an additional component, however, was made by EGRET in GRB 941017 \citep{Gonzalez-03}. 
In most cases, the additional power-law component extends to low energies ($\sim\,$few keV) and exceeds the Band component below a few tens of keV,  
forming a low-energy excess. At~high energies, this power-law component is detected up to $\sim10^{-0.5}$--$10^{1.5}$ GeV 
with photon index $\alpha_{\rm PL}\sim-1.9$ to $-1.5$. In~some cases (e.g., GRB 090926A and GRB 190114C), 
however, this component shows a high-energy turnover at early times before becoming a strict power-law as the spectral break 
moves above the LAT energy window. Example time-integrated and time-resolved spectra for the short-hard GRB 090510 
\citep{Ackermann+10} and long-soft GRB 090926A~\citep{Ackermann+11} are shown in Figure~\ref{fig:PL-Component}.

The low-energy excess presents a challenge for leptonic scenarios, e.g.,~SSC, that can only explain the excess emission above the peak of the sub-MeV emission 
and remain subdominant to the main synchrotron (Band) component at low energies. Such an excess can be produced in hadronic models featuring direct proton 
synchrotron emission 
\citep{Razzaque+10} or photohadronic cascades \citep{Asano+09a}. Theoretical modeling of the prompt spectrum of GRB 090510 using the latter model is shown in the 
bottom-left panel of Figure~\ref{fig:PL-Component} for two different ratios of $U_B/U_\gamma=\{10^{-3},10^{-1}\}$, 
where $U_B$ is the energy density of the comoving B-field, and $U_\gamma$ is the energy density of the Band component (shown with a 
black dashed curve; not produced by the same secondary pairs that IC scatter the Band component). In~this case, the~low- and high-energy excesses are given by synchrotron and inverse-Compton emissions from the secondary $e^\pm$-pair cascades (shown by thin red curves without absorption). The~peak around $\sim\,$GeV arises 
from absorption due to $\gamma\gamma$-annihilation.

Alternatively, models that attribute the origin of the additional power-law component to afterglow emission also find it challenging to explain the low-energy 
excess. As~mentioned earlier, in~simple external forward shock models, e.g., \cite{Kumar-Barniol-Duran-09,Kumar-Barniol-Duran-10}, the~dominant contribution 
from the afterglow occurs at $t=t_{\rm dec}\gtrsim t_{\rm GRB}$; therefore, these models cannot explain the origin of the additional power-law component in 
the prompt emission spectrum. In~the model put forth by \citet{Beloborodov+14}, the~inverse-Compton emission (as shown in the bottom-right panel of 
Figure~\ref{fig:PL-Component}) from shock-heated electrons behind the forward shock that sweeps up pre-accelerated and pair-loaded ISM remains sub-dominant at 
low-energies and~therefore cannot explain the low-energy~excess.

The low-energy excess is not always modeled as a low-energy extension of the power-law component that dominates 
at high energies above the Band component. In~some cases, it has been interpreted as a combination of a Band plus 
photospheric (quasi-thermal) components that jointly produce this excess, e.g., \cite{Guiriec+11,Guiriec+15a,Guiriec+16b}. This 
degeneracy produced by different spectral models describing the same data equally well further adds to the complexity of the 
underlying emission~mechanism.

\subsection{Long-Lived HE~Emission}
\label{sec:long-lived-emission}

\begin{figure}[H]
    \includegraphics[width=0.505\textwidth]{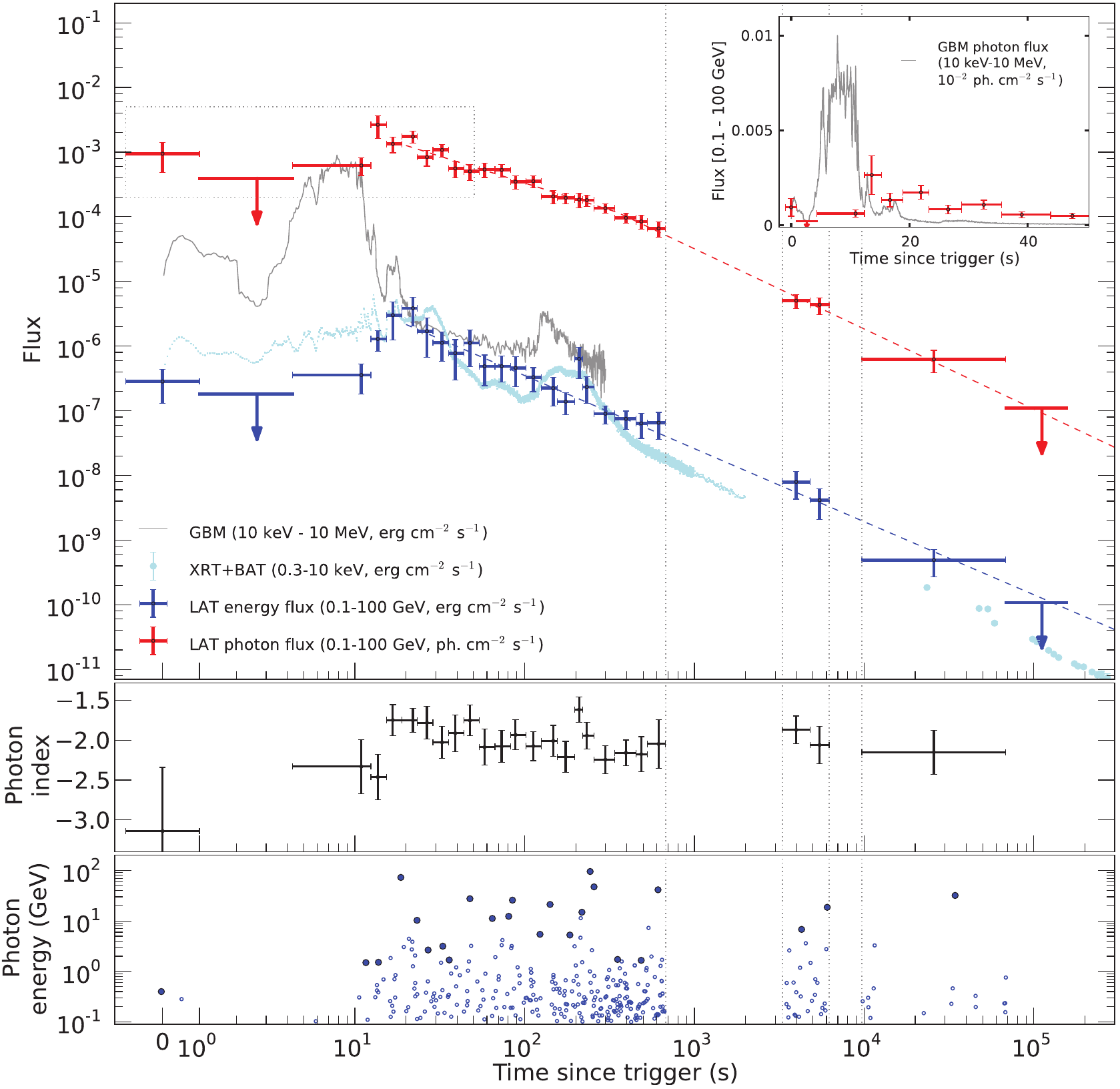}\quad\ \ \ 
    \includegraphics[width=0.377\textwidth]{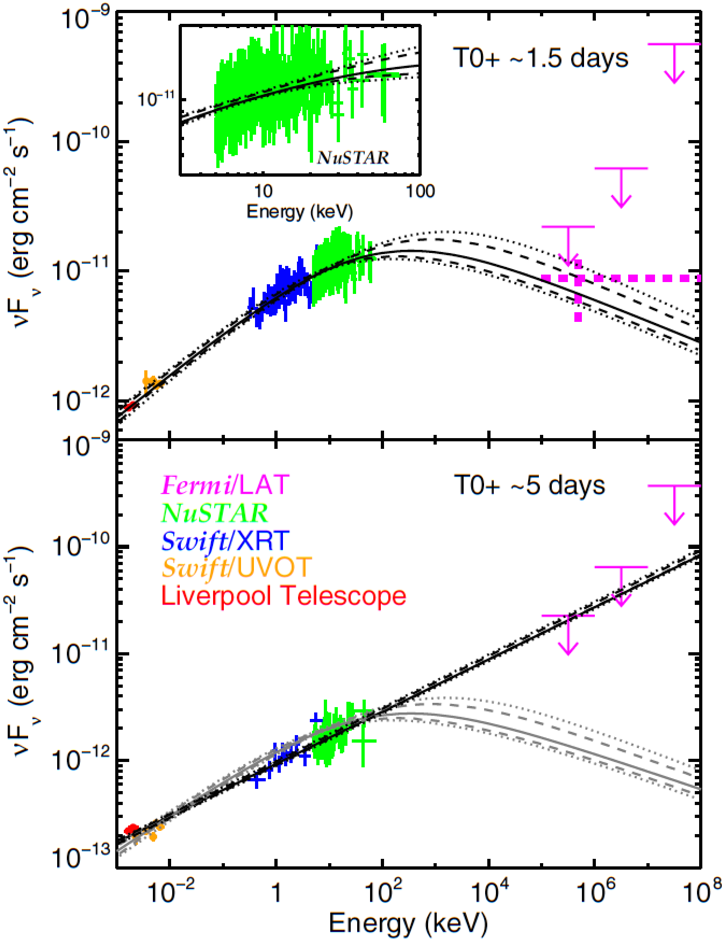}
    \vspace{0.25cm}
    \caption{\textbf{{Left}}: Temporally extended \textit{Fermi}-LAT emission ({from} \citep{Ackermann+14}, Reprinted with permission from AAAS). \textit{\textbf{Top}}: LAT energy flux (blue)
and photon flux (red) light curves. The~photon flux light curve shows a
significant break at a few hundred seconds (red dashed line), whereas the
energy flux light curve is well described by a single power law (blue dashed
line). The~10 keV to 10 MeV (GBM, gray) and 0.3 to 10 keV (XRT + BAT, light
blue) energy flux light curves are overplotted.
\textit{\textbf{{Middle}}}: LAT photon index. \textit{\textbf{{Bottom} 
}}: Energies of all the photons with probabilities
$>$90\% of being associated with the GRB. Solid circles correspond
to the photon with the highest energy for each time interval. The~vertical gray
lines indicate the first two time intervals during which the burst was occulted
by Earth. \textbf{Right}: A smoothly broken power-law 
synchrotron afterglow model \citep{Granot-Sari-02} fit to the optical to GeV spectrum of GRB\;130427A (from \citep{Kouveliotou+13}, \copyright AAS. Reproduced with permission). Broadband SEDs are shown during
the first (\textit{\textbf{{top}}} panel) and the second (\textit{\textbf{{bottom}}} panel) NuSTAR epochs. The~Fermi LAT upper limits are shown as arrows, and the extrapolation of the LAT flux
light curve is shown as a dashed magenta cross (only during the first epoch).
The second epoch (\textit{\textbf{bottom}} panel) is fit with a power law (black line); the fit to the first
epoch is scaled down and superposed on the second epoch data for comparison
(in gray).
    \label{fig:GRB130427A}
    }
\end{figure}

At an early time, when the sub-MeV prompt emission is still active, the~HE emission detected by the \textit{Fermi}-LAT shows 
significant temporal variability, which in many cases, e.g.,~\cite{Abdo+09a} is correlated with the sub-MeV emission. This 
can be attributed to the HE emission having originated in the same spatial region as the sub-MeV component. After~the prompt 
emission ceases and the afterglow commences, the~HE emission shows a temporally smooth and long-lasting decay with 
$\mean{d\log F_\nu/d\log t}\approx-1$ and a standard deviation of 0.8 (in some cases a broken power-law fit to the lightcurve 
is statistically preferred) \citep{Ajello+19}. This is often referred to as the \textit{LAT extended emission} 
(see Figure~\ref{fig:LAT-GRBs-Lightcurves} and left panel of Figure~\ref{fig:GRB130427A}). Since it lacks the short timescale 
variability and lasts much longer, it is naturally interpreted as the HE tail of the afterglow emission from the external forward~shock.

As discussed in Section~\ref{sec:Delayed-Onset}, synchrotron afterglow emission from an adiabatic \citep{Kumar-Barniol-Duran-09,Kumar-Barniol-Duran-10}  
(or possibly in some cases from a radiative \citep{Ghisellini+10}) blast wave can very well explain the temporal decay index of the LAT extended 
emission. The~agreement with 
multi-wavelength observations (see, e.g.,~\citep{Kumar-Barniol-Duran-10}) suggests that it certainly is a strong candidate for the late-time extended emission, 
even though this model may not be the correct description for the early time (with its delayed onset and low-energy spectral excess if produced by the same 
non-thermal component) LAT~emission. 

The one problem this scenario faces is the detection of (V)HE photons at late times. In~several GRBs, HE photons with observed energy $E\gtrsim10\,$GeV arrive at 
$t\sim10^2$--$10^3$~s, much after the cessation of the prompt GRB emission \citep{Ajello+19}. The~origin of such photons using the standard leptonic synchrotron 
afterglow scenario is difficult to explain as they violate $E_{\rm syn,max}$. According to this limit, to~produce photons with energy $(1+z)E\gtrsim10\,$GeV in 
the cosmological rest-frame of the source would require bulk $\Gamma>10^2$ at late times, which is nearly impossible. Therefore, our assumptions regarding particle 
acceleration at shock fronts must be revised (see further discussion in Section~\ref{sec:GeV-Afterglow-GRB130427A} below). The~alternative is SSC emission which would 
manifest as an additional spectral component in the LAT energy band and would also be detected at very high~energies.

\subsection{Constraints on Bulk $\Gamma$}
Since GRBs are extremely luminous sources, a~typical photon near the $\nu F_\nu$ peak with energy $E\sim E_{\rm pk}\sim m_ec^2$ would see a large optical depth 
$\tau_{\gamma\gamma}\gg1$ to pair production ($\gamma\gamma\to e^+e^-$) \citep{Piran-99}. For~a Newtonian source, this would imply a huge compactness 
$\ell\equiv\sigma_TU_\gamma R/m_ec^2$ (Thomson optical depth of pairs if all photons pair produce), where $U_\gamma$ is the radiation field energy density, which 
would result in a nearly blackbody spectrum, in~stark contrast with the observed prompt GRB non-thermal spectrum. The~solution to this so-called 
\textit{compactness problem}, is that the emission region must be moving towards us at ultra-relativistic speeds with bulk LF $\Gamma\gtrsim 10^2$ 
\citep{Goodman-86,Paczynski-86,Rees-Meszaros-92}. The~observed energy $E_{\rm cut}$ where the prompt GRB spectrum would display a cutoff due to 
$\gamma\gamma$-annihilation is sensitive to $\Gamma$; therefore, an~observation of such a cutoff yields a direct estimate of $\Gamma$, which is difficult to 
obtain otherwise. Spectral cutoffs have only been observed in a handful of GRBs, e.g.,~\citep{Ackermann+11,Tang+15,Vianello+18}, and in most cases, the 
spectrum above the $\nu F_\nu$ peak is a featureless power law extending to some $E_{\rm max}$, the~maximum photon energy detected by the instrument. In~such 
cases, only lower limits can be placed on the bulk LF with $\Gamma>\Gamma_{\rm min}$. The~maximum possible bulk LF for a given $E_{\rm cut}$ is given by 
$\Gamma_{\max}=(1+z)E_{\rm cut}/m_ec^2$, and~the true bulk LF is $\Gamma=\min[\Gamma_{\min},\Gamma_{\max}]$, e.g., \citep{Gill-Granot-18a}.

In several bright GRBs detected by \textit{Fermi}-LAT, estimates of $\Gamma_{\min}$ have been obtained for a given $E_{\max}$ using a simple one-zone 
analytic formalism, e.g., \citep{Woods-Loeb-95,Baring-Harding-97,Lithwick-Sari-01}, with~$\Gamma_{\min}\approx900$ for GRB 080916C \citep{Abdo+09c}, 
$\Gamma_{\min}\approx1200$ for GRB 090510 \citep{Ackermann+10}, and $\Gamma_{\min}\approx1000$ for GRB 090902B \citep{Abdo+09b}. When a more sophisticated 
formalism \citep{Granot+08,Zou+11,Hascoet+12} that includes temporal, spatial, and~angular dependence of the radiation field, and~which is verified by 
numerical simulations \citep{Gill-Granot-18a}, is applied, it yields a $\Gamma_{\min}$ estimate smaller by a factor of $\sim$2--3 (similar results were 
obtained by \citep{Li-10,Aoi+10}). In~GRBs that show a high-energy spectral cutoff, bulk LF of $\Gamma\sim$ 100--400 have been obtained using detailed numerical 
models \citep{Vianello+18}.


\section{High-Energy (GeV) Afterglow and GRB\;130427A}
\label{sec:GeV-Afterglow-GRB130427A}



GRB\;130427A was an exceptionally bright GRB \citep{Preece+14,Ackermann+14,Maselli+14} that occurred at a relatively small redshift of $z=0.340$ \citep{Levan+13}. Besides~being observed by the \textit{Fermi}-GBM \citep{vonKienlin13} and LAT \citep{Ackermann+14}, its extremely intense emission was also detected by other satellites (\mbox{AGILE \citep{Verrecchia+13}}, Konus-Wind  \citep{Golenetskii+13}, RHESSI \citep{Smith+13}, Swift \citep{Maselli+13}), which enabled multiple ground- and space-based follow-up observations. GRB\;130427A was a unique event in that it had the largest fluence ($\approx4.2\times10^{-3}\;\rm{erg}\;\rm{cm}^{-2}$), highest-energy photon at the time (95\;GeV), longest $\gamma$-ray duration (20 hours in the $0.1-100\;$GeV energy range), and~one of the largest isotropic energy releases ($E_{\gamma,\rm{iso}}\approx1.4\times10^{54}\;$erg) ever observed from a~GRB. 

In addition to its phenomenal prompt emission, the~afterglow emission GRB\;130427A provided extremely interesting physical insights. In~particular, as~described below, its
temporal and spectral analyses challenge the most widely accepted model for the afterglow phase of GRBs. Its prompt emission lasted about $T_{90}=276\pm5\;$s (at \mbox{15--150\;keV; \citep{Maselli+14})}, and~after it subsided, the observed emission was clearly dominated by the \linebreak{afterglow~\citep{Maselli+14,Perley+14,Kouveliotou+13}}, showing a smooth power-law flux decay as well as a typical afterglow-like spectrum (see Figure~\ref{fig:GRB130427A}). Moreover, while the reverse shock emission appears to dominate at early times at low frequencies, it does not dominate beyond the optical (even at early times), where the observed emission is dominated by the forward shock all along \citep{Perley+14}. The~values  of the temporal and spectral indices in the power-law segment $\nu_m<\nu<\nu_c$ (PLS G from \citep{Granot-Sari-02}) , $F_\nu\propto\nu^{-0.69\pm0.01}t^{-1.30\pm0.05}$, imply an external density profile $\rho_{\rm ext}\propto R^{-k}$ with \mbox{$k=1.4\pm0.2$ \citep{Kouveliotou+13}}, suggesting that the GRB progenitor star's wind mass loss rate to velocity ratio ($\dot{M}_w/v_w\propto R^{2-k}$ or $\dot{M}_w(\tilde{t})\propto v_w(\tilde{t})^{3-k}\tilde{t}^{\,2-k}$ where $\tilde{t}=R/v_w(\tilde{t})$ is the wind ejection time prior to the stellar explosion leading to the GRB) slightly decreased towards the end of its~life.

Most importantly, the~high-energy emission from the afterglow of GRB\;130427A was not only detected by the \textit{Fermi}-LAT for 20 hours (see \textit{top-left} panel of Figure~\ref{fig:GRB130427A}), but~also included multiple high-energy photons up to very late times (see \textit{bottom-left} panel of Figure~\ref{fig:GRB130427A}) that were clearly in excess of the maximum synchrotron photon energy, $E_{\rm syn,max}$. This upper limit on the energy of synchrotron photons is derived
by equating the electron acceleration and synchrotron radiative cooling timescales, assuming a single acceleration and emission
region \citep{Guilbert+83,deJager+96,Kirk-Reville-10,Piran-Nakar-10}. While there was some evidence of $E_{\rm syn,max}$ violation in previous \textit{Fermi}-LAT GRBs (e.g., \citep{Abdo+09c,Piran-Nakar-10}), in~those cases on the one hand the violation was weaker (by a smaller factor and with fewer photons of less statistical significance), and~on the other hand a different emission mechanism was a viable alternative explanation. In~GRB\;130427A, the~long-lasting
($\sim$1~day) \textit{Fermi}-LAT afterglow included a 32\;GeV photon
after 34\;ks, and~altogether five $>$ 30\;GeV photons after
$>$200\;s (with probability $>$ 99.9\% of being associated with GRB\;130427A). All five significantly exceed $E_{\rm syn,max}$, by~factors
of at least 6.25 for $k = 0$ and 9.20 for $k = 2$ (using Equation~(4)
of \citep{Piran-Nakar-10}). 

\textls[-15]{This has led to suggestions that the \textit{Fermi}-LAT high-energy photons were not synchrotron radiation, but~instead arose from a distinct high-energy spectral
\mbox{component \citep{Ackermann+14,Fan+13}}.  All such options require that the high-energy part of the \textit{Fermi}-LAT detected energy range (above a few to several 
GeV, depending on the exact time) should be dominated by a distinct spectral component, while lower energies are dominated by the usual afterglow synchrotron 
spectral component. Such an option was considered by \citep{Kouveliotou+13},  who fit the SED from optical to GeV at two epochs ($\sim$1.5 an 5 days) where 
observations were also performed by the Nuclear Spectroscopic Telescope ARray (NuSTAR) in the 3--79\;keV energy range (see \textit{right} panels of  Figure~\ref{fig:GRB130427A}). The~spectrum was fit to a detailed synchrotron afterglow model \citep{Granot-Sari-02}, which provided a good fit at both epochs. 
Moreover, at~the first and more constraining epoch (\mbox{$\sim$1.5 days}), the~\textit{Fermi}-LAT flux that is extrapolated by a factor of $\lesssim$2 in time agrees very 
well with this synchrotron afterglow model. Furthermore, both this, as~well as the simultaneous upper limit by the \textit{Fermi}-LAT \citep{Ackermann+14,Kouveliotou+13}, 
and more importantly the nearly simultaneous VHE upper limit by the Very Energetic Radiation Imaging Telescope Array System (VERITAS) 
\citep{Aliu+14}, hardly leaves any room for a distinct high-energy spectral~component.}

Therefore, this strongly suggests that the late-time Fermi LAT high-energy photons in GRB\;130427A are indeed afterglow synchrotron radiation. This provides 
the strongest direct observational support for a genuine violation of $E_{\rm syn,max}$ by synchrotron photons. As~the latter arise from the afterglow forward 
shock, this challenges our understanding of particle acceleration and magnetic field amplification in relativistic collisionless shocks. In~particular, at~
least one of the assumptions in the derivation of $E_{\rm syn,max}$ must be incorrect, requiring a modification of our understanding of afterglow shock~physics. 

While this potential problem was known before, these results from GRB\;130427A \citep{Kouveliotou+13} have made it much harder to circumvent (and the VHE 
upper limit by VERITAS played an important role). A~possible solution to this problem may lie in modifying the assumption of a single uniform region where 
both the acceleration of electrons and radiation from them occurs. Instead, one can allow for  a lower magnetic field acceleration region and a higher magnetic 
field synchrotron radiation region (e.g., \citep{Lyutikov10,Kumar+12}). Such a situation may arise for diffusive shock acceleration (Fermi Type I) if the 
tangled shock-amplified magnetic field decays on a short length scale behind the shock front. In~this case, most of the high-energy radiation is emitted just 
behind the shock from where the magnetic field has not decayed significantly, while the highest-energy electrons are accelerated further downstream where 
the magnetic field is lower \citep{Kumar+12}. This puzzle is still far from being resolved and poses a serious challenge to our understanding of relativistic 
collisionless shock~physics.

\section{Very-High-Energy (TeV) Afterglow}
\label{sec:TeV-Afterglow}
\textls[-15]{The detection by EGRET of MeV-GeV photons over $\sim$90~min from GRB \mbox{940217 \citep{Hurley+94}}, as~well as the hard additional spectral component in 
the prompt emission of GRB 941017~\citep{Gonzalez-03}, led to the consideration of SSC prompt \citep{Guetta-Granot-03b,Peer-Waxman-04,Gupta-Zhang-07,Bosnjak+09,Peer+12} 
and afterglow radiation~\citep{Panaitescu-Meszaros-98b,Dermer+00,Panaitescu-Kumar-00,Zhang-Meszaros-01,Sari-Esin-01} 
and searches for VHE TeV photons by ground-based detectors. TeV emission is expected to be observed only from relatively nearby GRBs due to absorption of VHE $\gamma$-rays by 
$\gamma\gamma$-annihilation on EBL photons from the more distant sources. The~Universe starts to become opaque to VHE photons with $E\gtrsim1\,$TeV for redshifts 
$z\gtrsim0.08$ \citep{Stecker+06}. Early efforts at detecting VHE radiation from GRBs were made by the Milagro instrument, an~extended air shower 
detector, and~hints of VHE photon ($3\sigma$) detection from GRB 970417A were found by Milagrito~\citep{Atkins+00} {(see \citep{Razzaque+09a} for prompt TeV $\gamma$-ray emission model for this detection)}, the~smaller and less sensitive prototype 
detector. Over~the last two decades, imaging atmospheric Cherenkov telescopes (IACTs), namely the Very Energetic Radiation Imaging Telescope Array System 
(VERITAS; \citep{Weekes+02}), the Major Atmospheric Gamma Imaging Cherenkov (MAGIC; \citep{MAGIC-COLLAB+05}), and~ the high-energy Stereoscopic System 
(H.E.S.S; \citep{HESS-COLLAB+04}) have been routinely monitoring for VHE radiation from~GRBs.}



\begin{figure}[H]
    \includegraphics[width=0.45\textwidth]{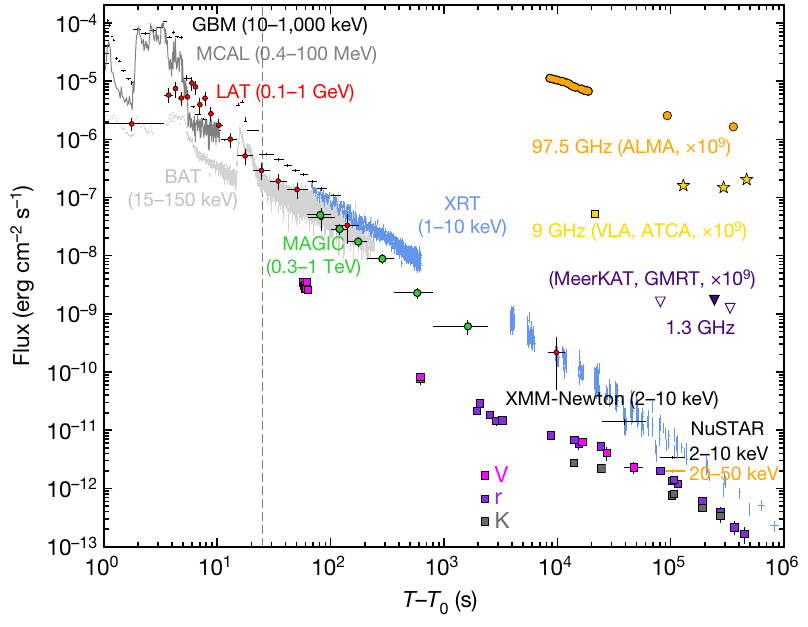}\quad\quad
    \includegraphics[width=0.45\textwidth]{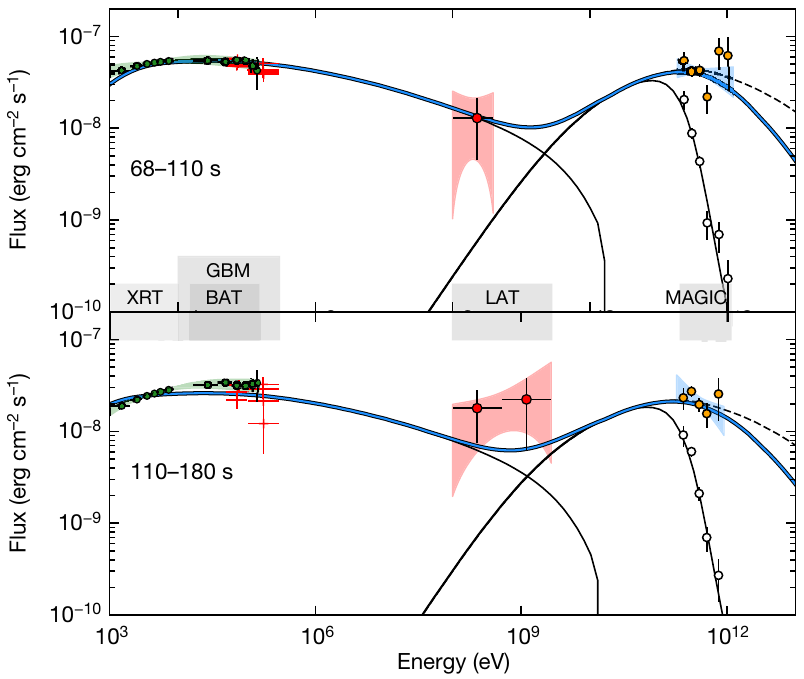}
    \caption{\textbf{Left}: {Multi-wavelength} 
     (Radio to TeV $\gamma$-rays) lightcurve of GRB 190114C. The~dashed verticle line indicates the time when the 
    highly variable prompt emission ended. 
    \textbf{Right}: Broadband high-energy spectrum of GRB 190114C at different time intervals along with the synchrotron + SSC model fits. The~shaded 
    contours shown in different colors indicate the $1\sigma$ uncertainties on the model fit to the data. MAGIC data points shown in yellow have been 
    corrected for the EBL absorption, whereas the open white circles are the actual observations. The~black dashed curve shows the SSC spectral component without 
    attenuation from internal $\gamma\gamma$ and EBL absorption. Figures {from} \citep{MAGIC_GRB190114C} (Reprinted by permission from Springer Nature Customer Service Center GmbH: Springer Nature, 
    Observation of inverse Compton emission from a long $\gamma$-ray burst, MAGIC Collaboration, \copyright 2019). }
    \label{fig:GRB190114C-lightcurve-spectrum}
\end{figure}

These efforts bore fruit in January of 2019 when MAGIC announced the ($\gtrsim$50$\sigma$) detection of VHE ($\sim$0.2--1~TeV) 
photons from GRB 190114C \citep{MAGIC-Collab-GRB190114C}. Figure~\ref{fig:GRB190114C-lightcurve-spectrum} shows the multi-wavelength lightcurve 
and broadband afterglow spectrum of this burst. 
GRB 190114C had a redshift of $z=0.424$, and its prompt emission was detected by several space-based $\gamma$-ray instruments \citep{Ajello+20,Ursi+20} that 
measured an isotropic-equivalent ($1\,$keV to $10\,$GeV) energy release of $E_{\rm iso}\simeq3\times10^{53}\,$erg over a duration of $\sim$25~s (shown 
by the dashed vertical line in Figure~\ref{fig:GRB190114C-lightcurve-spectrum}). MAGIC detected afterglow VHE $\gamma$-ray photons 
from $\sim$60~s to $\sim$2400~s and measured an $E_{\rm iso,TeV}\simeq4\times10^{51}\,$erg, which is only a lower limit due to the late start of observations 
and could be as high as $\sim$10\% of the energy released in softer $\gamma$-rays. This was the first time that time-resolved afterglow spectra all the way 
up to TeV energies were obtained in any GRB observed to date. This naturally has important implications for GRB afterglow physics and overall energetics of the 
system.

After GRB 190114C, a~few other GRBs (160821B \citep{Acciari+21}with a low detection significance of $\sim$3$\sigma$, 180720B \citep{Abdall+19}, 
190829A \citep{Abdalla+21}, 201216C \citep{Fukami+21}) were reported to have been detected at sub-TeV to TeV energies by both H.E.S.S and MAGIC 
(see the reviews by \citet{Nava-21,Noda-Parsons-22} for more details). 

\subsection{Key Results \and Implications}
In the following, we briefly discuss the most important implications for GRB physics from the detection and theoretical modeling of $\sim$ TeV afterglow~emission.

\subsubsection{{IC~Emission} Is Needed to Explain the VHE $\gamma$-Rays}
As shown in the right panel of Figure~\ref{fig:GRB190114C-lightcurve-spectrum}, the~hardening of the MAGIC-detected VHE spectrum with 
respect to the LAT detected HE spectrum in GRB 190114C indicates the presence of an additional spectral component. It simply cannot 
be explained with synchrotron emission from the external forward shock alone. Several works that use analytical/semi-analytical \citep{Wang+19,Zhang+20,Joshi-Razzaque-21,Yamasaki-Piran-21,Jacovich+21} and numerical models \citep{Asano+20,Derishev-Piran-21} have 
now been devoted to explaining the $\sim\,$TeV emission as SSC {or a combination of EIC and SSC \citep{Zhang+21}}.  
Significant differences between the (semi-)analytical and numerical models arise due to inclusion and more accurate handling of 
some of the non-linear processes, such as pair cascades due to internal $\gamma\gamma$ absorption and KN effects. In~the end, the~
obtained shock microphysical parameters indicate that these bursts are not very different from the ones that are not detected with 
a VHE component, which may suggest that SSC afterglow emission is rather common. In~that case, it becomes important to take into 
account the energy radiated in the SSC component to understand the global energetics of the bursts. For~example, the~energy in the SSC 
component was $\sim$40\% of that radiated in the main synchrotron afterglow component for GRB 190114C \citep{MAGIC-Collab-GRB190114C}.
Similar inferences regarding the total energy budget were also drawn before and around the first GeV detections from GRBs by the 
\textit{Fermi}-LAT. It was later shown that on average $E_{\rm GeV}/E_{\rm MeV}\lesssim0.1$, and at best the two become comparable for 
rare individual LAT GRBs. In~GRB\,190114C the detected TeV emission is from the afterglow. Therefore, it does not affect the prompt GRB energy~budget. 

{
An alternative to IC emission that can explain the VHE TeV $\gamma$-rays is photohadronic emission, as~demonstrated by 
\citet{Sahu-Fortin-20,Sahu+22}. In~their model, VHE $\gamma$-ray photons are produced via the $p\gamma\to\Delta^+$ process, 
which then produces neutral pions that decay into $\gamma$-ray photons (see Section~\ref{sec:pion-decay}). The~seed photons that 
interact with the protons can be of synchrotron or SSC origin as produced in the afterglow forward shock. 
}

\subsubsection{Constraints on Shock Microphysical~Parameters}
When fitting afterglow observations, the parameter space is usually degenerate, and unique values of the shock 
microphysical parameters cannot be obtained. The~parameter space consists of $E_{\rm k,iso}$, the~
isotropic-equivalent total kinetic energy of the flow, $n$, the~number density of the ISM or its normalization 
($n\propto AR^{-2}$) for a wind circumburst medium, and~the shock microphysical parameters $\epsilon_e$, 
$\epsilon_B$, and~$\xi_e$. The~power-law index $p$ of the particle energy distribution is uniquely determined 
from the broadband spectrum. It is generally assumed that $\xi_e=1$, in~which case the remaining four afterglow 
model parameters $(E_{\rm k,iso}^*,~n^*,~\epsilon_e^*,~\epsilon_B^*)$ can be uniquely determined using 
$(\nu_a,\nu_m,\nu_c,F_{\nu,\max})$, leaving the degeneracy due to $\xi_e$ \citep{Eichler-Waxman-05} where 
all values of $(E_{\rm k,iso},~n,~\epsilon_e,~\epsilon_B)=(\xi_e^{-1}E_{\rm k,iso}^*,~\xi_e^{-1}n^*,~\xi_e\epsilon_e^*,~\xi_e\epsilon_B^*)$ 
fit the data equally well for any $m_e/m_p<\xi_e\leq1$. This degeneracy may possibly be broken when accounting 
for the emission, absorption, or propagation effects of the thermal electrons \citep{Sagiv+04,Toma+08,Giannios-Spitkovsky-09,Ressler-Laskar-17,Warren+22}. 
{In addition to these parameters, multi-wavelength modeling of TeV afterglows can potentially be used to 
constrain $\gamma_M$ and in turn the acceleration efficiency ($\kappa_{\rm acc}$) of relativistic collisionless shocks. 
Such constraints can then be used for comparison with first principles PIC simulations of Weibel-mediated collisonless shocks 
that do predict the value of $\gamma_M$.}



In Table~\ref{tab:TeV-GRBs}, we list the afterglow fit parameters of GRBs that were detected at very high energies. In~all cases, 
the energy 
deposited in power-law electrons is much larger than that in the shock-generated B-field, i.e.,~$\epsilon_B\ll\epsilon_e$. Consequently, 
this yields the Compton-$y$ parameter larger than unity which results in producing a bright SSC component detected by MAGIC and 
H.E.S.S. In~most works, the afterglow shock microphysical parameters are taken to be constant throughout the afterglow evolution; 
however, \citet{Misra+21} report the possibility of evolving microphysical parameters to explain the long-term radio/mm afterglow 
of GRB~190114C.
\begin{table}[H] 
\caption{Afterglow fit parameters for GRBs with VHE emission. {$^\dagger~$Shows $A_\star$ parameter for a \mbox{wind~environment.}} \label{tab:TeV-GRBs}}
\setlength{\tabcolsep}{2.4mm}
\begin{tabular}{ccccccccc}
\toprule
\textbf{GRB} & \boldmath{$E_{\gamma,\rm iso,53}$} & \boldmath{$E_{\rm k,iso,53}$} & \boldmath{$n\,({\rm cm}^{-3})$} & \boldmath{$p$} & \boldmath{$\epsilon_{e,-2}$} & \boldmath{$\epsilon_{B,-5}$} & \boldmath{$\xi_e$} & \textbf{Ref.}\\
\midrule
180720B	& 6 & $10$ & 0.1 & 2.4 & 10 & 10 & 1 & \citep{Wang+19} \\
 & & $\sim$40 & 1 & 2.5 & 1 & 5 & 1 & \citep{Fraija+19} \\
190114C	& 2.5 & 8 & 0.5 & 2.6 & 7 & 8 & 1 & \citep{MAGIC_GRB190114C}\\
	& & $10$ & 1.0 & 2.3 & 6 & 90 & 0.3 & \citep{Asano+20}\\
	& & $3$ & 2 & 2.5 & $\sim$10 & $\sim$400 & 1 & \citep{Derishev-Piran-21}\\
	& & $40$ & $^\dagger~2\times10^{-2}$ & $2.18$ & $3.3$ & $1200$ & 1 & \citep{Joshi-Razzaque-21}\\
190829A	& $2\times10^{-3}$ & $\sim$2.5 & 0.21 & 2.0 & $\sim$3 & $\sim$2.5 & $<$0.065 & \citep{Salafia+21} \\
	& & $0.098$ & 0.09 & 2.2 & 39 & 8.7 & $0.34$ & \citep{Zhang+21} \\
	& & $0.3$ & $0.01-0.1$ & 2.5 & 10 & 10 & $1$ & \citep{Dichiara+21} \\
\bottomrule
\end{tabular}
\end{table}

\subsubsection{VHE $\gamma$-Rays as Electromagnetic Counterparts of Binary NS~Mergers}

{
The detection of afterglow TeV $\gamma$-rays in long-soft GRBs has opened up the prospect of also detecting VHE emission in short-hard GRBs. 
Electromagnetic emission coincident with GWs was first detected from the binary NS merger in GW\,170817/GRB\,170817A. An~impressive 
multiwavelength follow-up by a number of ground- and space-based observatories tracked its peculiar afterglow emission. No $\sim$TeV 
$\gamma$-rays were detected \citep{Abdalla+17,Galvan+19,Salafia+21} for this relatively nearby ($\sim$40~Mpc) event as the relativistic 
jet was observed off-axis, and~none have been detected from other short-hard GRBs. Short-hard GRBs are detected more numerously at redshifts 
$z<1$ with a mean redshift of $\langle z\rangle\approx0.5$ 
in comparison to $\langle z\rangle\approx2$ for the long-soft GRBs. Therefore, attenuation of VHE $\gamma$-rays by the EBL is not as 
extreme for the short-hard GRBs (e.g., MAGIC detected GRB\,190114C had a redshift of $z\approx0.42$) as it is for the more distant 
long-soft GRBs. Catching the VHE emission in time from short-hard GRBs will require high sensitivity (due to lower fluences), 
shorter telescope slew times, as~well as a large field-of-view (since catching the prompt emission would require the GRB to be in the 
field-of-view without slewing). Fast follow-ups of GW triggers by existing (MAGIC, H.E.S.S, VERITAS, HAWC) 
and upcoming observatories, namely the Cherenkov Telescope Array (CTA; \citep{Acharya+13}), will play a pivotal role in the next several years. 
}

\section{Studying Non-GRB~Physics}
\label{sec:Non-GRB-Physics}

Since GRBs are the most electromagnetically luminous events in the Universe, they are observed up to cosmological 
distances. They emit HE and VHE photons that travel over cosmological distances on the way from the source to us. 
This can naturally be used to probe various processes involving these photons that may occur along their way to 
our detectors, which provide unique and valuable information about cosmology or basic~physics.  

\subsection{Constraining the Extragalactic Background Light Models with~GRBs}



The detection of HE photons from distant GRBs proves to be an excellent probe of the extragalactic background 
light (EBL) \citep{Knieske+04,Stecker+06,Franceschini+08,Razzaque+09b,Gilmore+09,Finke+10,Abdo+10a,Dominguez+11}, which is the cumulative star light 
emitted in the UV/optical to infrared energy range, i.e.,~$\sim$10$^{-3}$~eV to $10\,$eV ($\sim$0.1~$\upmu$m to 10$^3~\upmu$m), by~all the 
stars in the Universe. The~EBL is difficult to constrain otherwise due to contamination by the zodiacal and 
Galactic foreground light \citep{Hauser-Dwek-01}. After~the cosmic microwave background, the~EBL is the second dominant component that 
contributes to the diffuse radiation that pervades entire space. Star light with wavelength $\lesssim2\upmu$m is 
highly absorbed by dust in the host galaxy, with~only a fraction that escapes and contributes to the EBL. On the other hand, the dust re-radiates the star light but adds to the EBL in the infrared. As~discussed in 
Section~\ref{sec:Pair-Echoes}, VHE $\gamma$-ray photons with $E_\gamma\gtrsim1\,$TeV interact with the infrared 
background and produce $e^\pm$-pairs, while lower energy photons typically interact with UV/optical/NIR photons 
emitted directly by stars. The~attenuation of the HE spectra of TeV sources, including GRBs, caused by this 
effect can be used to constrain models of EBL, with~the assumption that the intrinsic spectrum can be extrapolated to 
higher energies using the lower energy part of the spectrum that is not affected by such~attenuation.

\begin{figure}[H]
    \includegraphics[width=0.48\textwidth]{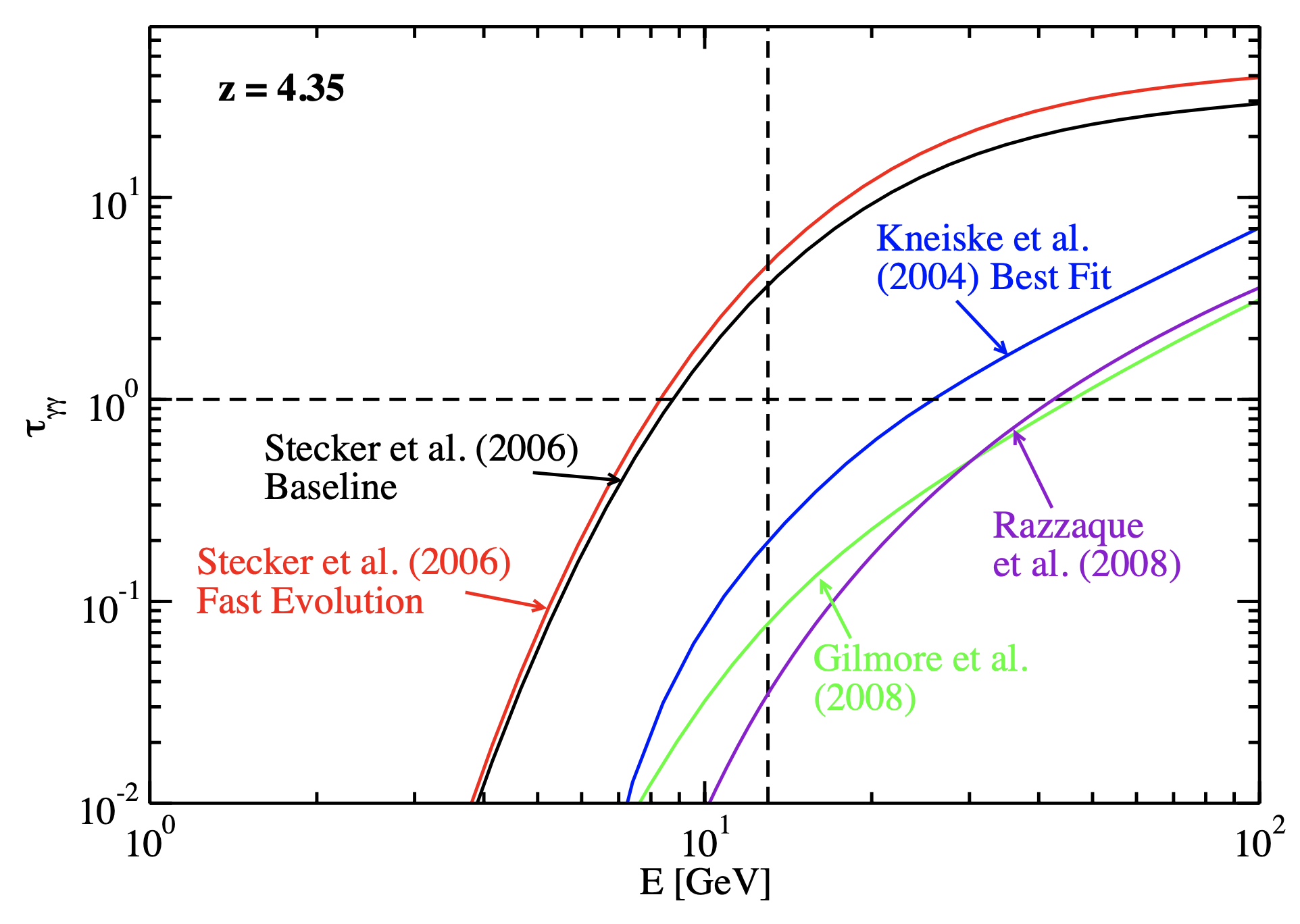}
    \caption{{Opacity} of the Universe to HE photons, emitted at a redshift of $z=4.35$, due to their interaction with 
    EBL photons as calculated using different models. The~vertical dashed line marks a $13.2\,$GeV photon detected 
    from GRB\,080916C. Figure adapted {from} \citep{Abdo+09c} (Reprinted with permission from AAAS).}
    \label{fig:EBL-Models}
\end{figure}

Before the  MAGIC detection of $\sim\,$TeV $\gamma$-rays in GRB\,190114C, constraints on EBL models were placed 
using the observations of HE photons from only a few GRBs. For~example, the~\textit{fast evolution} and \textit{baseline} 
models of \citet{Stecker+06} were disfavored at the $>$3$\sigma$ level by the detection of a $33.4\,$GeV photon from 
GRB\,090902B which was at a redshift of $z=1.822$ \citep{Abdo+09b}. Observations of higher redshift GRBs at high 
energies offer a better chance of constraining EBL models. A~$13.2\,$GeV photon was detected from GRB\,080916C 
having a redshift of $z=4.35$. The~opacity of the Universe to such a photon is shown in Figure~\ref{fig:EBL-Models} 
that compares different EBL models. The~suppression of the $\sim\,$TeV spectrum of a GRB due to the EBL was first clearly 
seen in GRB\,190114C \citep{MAGIC_GRB190114C,MAGIC-Collab-GRB190114C}, as~shown in Figure~\ref{fig:GRB190114C-lightcurve-spectrum}.

\subsection{Probing the Intergalactic Medium~B-Field}
\label{sec:IGMF-Constraints}
The space between galaxies is expected to be permeated by a very weak ($B_{\rm IGMF}\gtrsim10^{-20}\,$G) magnetic field 
\citep{Kronberg-94,Grasso-Rubinstein-01,Widrow-02,Kulsrud-Zweibel-08}, 
the origin, strength, and~coherence length of which are poorly understood. This intergalactic magnetic field (IGMF) 
possibly acted as the \textit{seed} magnetic field in galaxies and galaxy clusters, which was amplified to typical strengths of $\sim\,$$\upmu$G by a dynamo 
mechanism as well as flux conserving collapse during their formation. Therefore, its 
origin predates structure formation in the Universe. Since such field amplification processes are absent in the voids 
between galaxies, which would have otherwise erased the initial magnetic field properties, study of the IGMF can 
provide important insights into the origin of the seed field in galaxies, and it can be used to constrain physical 
processes in the early Universe that may have generated it. Contamination of the primordial IGMF is possible via 
magnetized outflows from active galaxies and galactic winds driven by star~formation.

One of the ways to study the IGMF, albeit indirectly, is by detecting $\sim\,$GeV \textit{pair echos} (see Section~\ref{sec:Pair-Echoes}) 
created by the IC scattering of CMB photons by $e^\pm$ pairs that in turn are produced by $\sim$TeV $\gamma$-rays (from an astrophysical source) 
annihilating with EBL \mbox{photons \citep{Plaga-95,Tavecchio+10}}. This causes several observable effects, e.g.,~time delays between the TeV 
and GeV signal \citep{Dai-Lu-02,Razzaque+04b,Murase+08} and extended $\gamma$-ray haloes \citep{Aharonian+94}, that can be 
used to constrain the properties of the IGMF. Non-observation of GeV $\gamma$-rays from persistent sources, e.g.,~TeV blazars, 
were used to derive a lower limit of $B_{\rm IGMF}\gtrsim10^{-16}\,$G for a coherence length of 10\,kpc to 1\,Mpc 
\citep{Neronov-Vovk-10,Ackermann+18}. 
The potential problem with persistent sources, when GeV radiation is observed, is that the pair echo photons can overlap with 
the intrinsic emission causing contamination. In~that regard, GRBs serve as better probes since there could be a clear 
temporal separation between the short-lived prompt emission and the detection of the longer-lived GeV pair~echo.

Before the detection of $\sim\,$TeV $\gamma$-rays from GRB\,190114C, constraints on the IGMF were obtained using 
GRB\,130427A from (i) the VERITAS upper limits at 100\,GeV at 0.82\,days, (ii)~\textit{Fermi}-LAT detection of a 32\,GeV photon 
at 34.4\,ks post-trigger, that could not be explained as synchrotron afterglow radiation, and~(iii) \textit{Fermi}-LAT upper 
limits at GeV energies at late times \citep{Veres+17}. From~the non-detection of GeV emission in 
\textit{Fermi}-LAT observations of the location of GRB\,190114C over a period up to 3 months, \citet{Wang+20} derive a lower limit 
of $B_{\rm IGMF}\approx2\times10^{-20}(\lambda_B/0.1\,{\rm Mpc})^{-1/2}\,{\rm G}$ for the coherence length of $\lambda_B<0.1\,$Mpc. 
On the other hand, \citet{Dzhatdoev+20} argued that the \textit{Fermi}-LAT flux upper limits were insufficient in constraining 
the IGMF and that the results of \citet{Wang+20} are in error due to overestimation of the pair echo~intensity.

\subsection{Lorentz Invariance~Constraints}
\label{sec:LIV-Constraints}

One of the tenets of the special theory of relativity is that the speed of light is a Lorentz invariant, i.e.,~it 
is the same in two Lorentz frames regardless of their relative motion. In~particular, it is independent of the photon 
energy (or wavelength), which is frame-dependent. The~underlying assumption is that this remains true at all length 
scales (or wavelengths) in nature, no matter how small. However, quantum effects are expected to strongly affect the 
nature of space-time at the Planck scale, corresponding to a length scale of $l_{\rm Planck}=\sqrt{\hbar G/c^3}\approx1.62\times10^{-33}\,$cm, 
or equivalently at an energy scale of $E_{\rm Planck}=M_{\rm Planck}c^2=\sqrt{\hbar c^5/G}\approx1.22\times10^{19}\,$GeV. 
Such effects can possibly lead to Lorentz invariance violation (LIV) where the speed of light changes with photon energy~\citep{Amelino-Camelia+98,Mattingly-05,Kostelecky+08}. In~some theories of quantum gravity, this effect can lead to 
dispersion as the photon propagates in the vacuum of space, such that its speed varies as $v_\gamma \approx c(1\pm E_\gamma/E_{\rm QG})$ 
for $E_\gamma\ll E_{\rm QG}$, where $E_\gamma$ is the photon energy, $E_{\rm QG}$ is the quantum gravity 
energy scale (expected to be $\sim$\,$E_{\rm Planck}$), and~the sign ambiguity depends on the dynamical framework. 
Since $E_{\rm QG}$ is much larger than observed photon energies, the~change in velocity is rather minute. However, 
this effect can accumulate over cosmological distance scales $D$, which makes GRBs ideal probes of~LIV.

This LIV effect would manifest as an arrival time difference between photons having different energies, with~
$\Delta t_{\rm LIV}\approx\pm(\Delta E_\gamma/E_{\rm QG})D/c$ and $\Delta E_\gamma=E_{\gamma,\rm high}-E_{\gamma,\rm low}$. 
Since GRBs show temporal variability as short as $\Delta t\sim\,$ms, their large 
distances $D\sim10^{28}L_{28}\,$cm can probe quantum gravity energy scales approaching the Planck scale using $\sim\,$GeV photons, 
$E_{\rm QG}\approx3\times10^{19}(\Delta E_\gamma/{\rm GeV})D_{28}\Delta t_{-2}^{-1}\,{\rm GeV}\sim E_{\rm Planck}$. 
This technique was employed in the case of short-hard GRB\,090510 that emitted a 31\,GeV photon 0.829\,s after 
the burst onset and which coincided in time with the last of the seven pulses comprising the prompt emission~\citep{Abdo+09a}. 
By using an unbinned analysis, in~both energy and time, testing different dispersion coefficients that would yield 
velocity differences of $\Delta v\approx E_{\gamma}/E_{\rm QG}$, and~then maximizing $\Delta v$ so that it yields the sharpest 
lightcurve, \citet{Abdo+09a} obtained a lower limit of $\vert\Delta t/\Delta E_\gamma\vert < 30\,{\rm ms\,GeV}^{-1}$ 
(at the 99\% confidence level) or equivalently $E_{\rm QG} > 1.2E_{\rm Planck}$. Using the same data for the short-hard 
GRB\,090510, this limit was somewhat improved using a more refined analysis \citep{Vasileiou+13}, and~a Planck-scale limit 
was also derived on space-time fuzziness and stochastic LIV \citep{Vasileiou+15}, which are motivated by the notion of space-time~foam.

\section{Outstanding~Questions}
\label{sec:Outstanding-Qs}
Even with only a handful of detections in the VHE domain, new questions have emerged. We briefly highlight some 
of the fundamental questions that may be resolved with future TeV detections as well as improved modeling of 
radiation~processes.

\begin{enumerate}

\item[(a)] \textit{{What makes GRBs TeV bright?} 
}: All TeV bright GRBs are also very bright in prompt $\gamma$-rays as well as in their X-ray 
afterglow emission. In~fact, apart from GRB\,190829A, the~rest of the TeV bright GRBs have high prompt $\gamma$-ray fluences that put them 
among the top 1\%, see Figure~1 of~\citep{Noda-Parsons-22}, as~is also evident from their high $E_{\gamma,\rm iso}\gtrsim10^{53}\,$erg from  
Table\,\ref{tab:TeV-GRBs}. Although~not all MeV-bright GRBs were observed at TeV energies, it begs the question why no TeV emission was 
detected from, e.g.,~130427A (one of the most energetic GRBs with $E_{\gamma,\rm{iso}}\approx1.4\times10^{54}\;$erg) by VERITAS and HAWC, and~
whether we would have seen TeV $\gamma$-rays from all such GRBs. The majority of the highly energetic GRBs are also more distant with $z>1.0$ 
see, e.g.,~Figure~3 of~\citep{Noda-Parsons-22}, 
which makes it challenging to detect their TeV emission due to suppression by $\gamma\gamma$-annihilation on EBL photons. Internal absorption 
due to $\gamma\gamma$-annihilation of IC photons (that produce the TeV component) on the seed synchrotron photons (that produce the X-ray afterglow) 
can also become important \citep{Derishev-Piran-21} and may perhaps be enough in some bursts to significantly suppress the VHE emission. 
Detailed semi-analytic numerical models including the effects of pair cascades and Klein--Nishina suppression that can explain the multi-wavelength 
spectral and temporal evolution may shed more light on the properties of the emission~region.

\item[(b)] \textit{{What causes the delayed onset of the \textit{Fermi}-LAT emission?}}:
The delayed onset of the HE emission w.r.t the $\sim\,$MeV prompt $\gamma$-rays as seen by the \textit{Fermi}-LAT has been interpreted as 
the peak of the standard afterglow emission \citep{Kumar-Barniol-Duran-09,Kumar-Barniol-Duran-10}, IC GeV flash from the pre-accelerated pair-rich 
circumburst medium swept up by the external forward shock \citep{Beloborodov-05b,Beloborodov+14}, acceleration time of protons in hadronic 
emission scenarios \citep{Razzaque+10}, and~the timescale over which the SSC radiation field builds up \citep{Asano-Meszaros-12}. The~latter two 
scenarios have some difficulties with, respectively, the~global energetics and limitation on the delay duration, and~the former two struggle with 
producing the observed variability at early times (while the sub-MeV prompt emission is ongoing) as both invoke emission from the external 
forward shock. Future and more sensitive observations of such delayed emission will be important in distinguishing between the different~models. 

\item[(c)] \textls[-5]{\textit{{What mechanism produces the \textit{Fermi}-LAT extended emission?}}: The smooth temporal decay of the GeV \textit{Fermi}-LAT 
extended emission naturally favors its origin in emission arising from the external forward shock. The~main question here is  whether 
the emission is entirely synchrotron radiation from non-thermal shock-heated \mbox{electrons \citep{Kumar-Barniol-Duran-09,Kumar-Barniol-Duran-10,Ghisellini+10}}, the~standard 
scenario, or~IC radiation from mostly quasi-thermal electrons as the blast wave encounters pair-rich and pre-accelerated circumburst medium 
\citep{Beloborodov-05b,Beloborodov+14}. The~latter scenario can only operate as long as there are softer seed photons that can Compton cool the thermal 
electrons. At~early times, they are the prompt sub-MeV photons that overlap the afterglow shock, and at later times softer photons can be of 
synchrotron origin \citep{Vurm+14}. Detailed numerical models of blast waves propagating into pair-enriched media and the comparison of afterglow lightcurves with 
observations over the entire duration of the LAT extended emission can shed more light on this~issue. }

\item[(d)] \textit{{What mechanism produces multi-GeV photons at late times?}}: The detection of $\gtrsim$10~GeV photons in several GRBs at late 
times ($t\sim10^2$--$10^3$~s) is puzzling. When their origin is interpreted as the standard afterglow synchrotron emission from shock-heated electrons, 
for which strong evidence came from the broad band (optical to GeV) SED fits of the afterglow of GRB\,130427A \citep{Kouveliotou+13}, it challenges 
our understanding of particle acceleration at relativistic collisionless shocks since the photon energy clearly violates $E_{\rm syn,max}$. 
{The alternative is IC (either SSC or EIC) afterglow emission, which can produce HE photons at late times. A~prime example is GRB\,190114C from 
which HE and VHE photons were detected by the \textit{Fermi}-LAT ($t\lesssim150\,$s) and MAGIC ($t\lesssim2400\,$s), respectively, at~late times.} 
Future such events with multi-wavelength constraints, especially at VHEs, along with numerical simulations of particle acceleration at shock fronts 
will be able to shed more light on this~issue.

\end{enumerate}

To answer the above questions, both leptonic and hadronic models have been discussed. Important constraints on the latter scenario are 
offered by multi-messenger observations that include follow-up and monitoring of GRBs by neutrino detectors \citep{Dorner+12}. Even non-detections offer 
very useful information about the underlying radiation mechanism. However, the~prospect of turning these non-detections into detections, or~at least 
providing more stringent upper limits, is looking better with the installation of km$^3$-scaled neutrino detectors, namely Baikal-GVD 
\citep{Avrorin+19} and KM3NeT~\cite{Adrian-Martinez+16}, in~the next few to several~years.

\section{Closing~Remarks}
\label{sec:Closing-Remarks}
The detection of afterglow TeV $\gamma$-rays in a few GRBs, first reported for GRB\,190114C, has opened up a new window for 
understanding the properties of relativistic collisionless shocks and radiation processes that operate near the shock fronts. 
VHE emission was anticipated in GRBs for some time, but~it remained undetected for decades, garnering only upper limits from 
ground-based imaging atmospheric Cherenkov detectors. Detailed spectral modeling of the afterglow TeV emission is now shedding new light 
on the global energetics of the system leading to better constraints on the prompt $\gamma$-ray emission efficiency. Moreover, 
detection of TeV emission during the prompt-GRB phase would help pin down its illusive emission mechanism(s). In~most 
cases, one-zone SSC emission is the most favored radiation mechanism for producing afterglow TeV photons, however, with~only a 
few sources the details of SSC emission from shock-heated relativistic electrons (or $e^\pm$-pairs) are not entirely clear. Future, 
{multi-messenger} and perhaps more sensitive, observations from low redshift GRBs will offer better opportunities to constrain 
microphysical processes at shock~fronts.

\vspace{6pt} 



 \authorcontributions{{Writing--original draft preparation--review and editing, R.G. and J.G. All authors have read and agreed to the published version of the manuscript.}}

\funding{{This research was funded in part by the ISF-NSFC joint research program under grant no. 3296/19 (J.G.).}}

 \dataavailability{Not applicable.} 

\acknowledgments{We are grateful to Kohta Murase and Soebur Razzaque for their feedback and comments that improved the quality of this article. We also thank the anonymous referees for their~comments. 
}

\conflictsofinterest{The authors declare no conflict of interest.
} 





\appendixtitles{no} 




\begin{adjustwidth}{-\extralength}{0cm}
\reftitle{References}



\end{adjustwidth}


%


\end{document}